\documentclass{elsarticle}
\usepackage[table]{xcolor}
\usepackage{indentfirst}
\usepackage{tkz-kiviat} 
\usepackage{booktabs}
\usepackage{chngpage}
\usepackage{url}
\usepackage{subfig}
\usepackage{enumitem}
\usepackage[markup=underlined]{changes}
\usepackage{threeparttable}
\usepackage{appendix}
\usepackage{adjustbox}
\usepackage{tikz}
\usetikzlibrary{calc,shadings,patterns,tikzmark}

\newcommand*{\hatchB}[1]{
\begin{adjustbox}{valign=t}
\begin{tikzpicture}[every node/.style={inner sep=0,outer sep=0}]
\draw[pattern color=green!70, pattern=vertical lines, draw=none] (0,0) rectangle (0.35,0.35) node[pos=.5] {#1};
\end{tikzpicture}
\end{adjustbox}}

\newcommand*{\hatchW}[1]{
\begin{adjustbox}{valign=t}
\begin{tikzpicture}[every node/.style={inner sep=0,outer sep=0}]
\draw[pattern color=red!70, pattern=horizontal lines, draw=none] (0,0) rectangle (0.35,0.35) node[pos=.5] {#1};
\end{tikzpicture}
\end{adjustbox}}

\newcommand*{\hatchE}[1]{
\begin{adjustbox}{valign=t}
\begin{tikzpicture}[every node/.style={inner sep=0,outer sep=0}]
\draw[draw=none] (0,0) rectangle (0.35,0.35) node[pos=.5] {#1};
\end{tikzpicture}
\end{adjustbox}}

 \newcommand{\sig}{\cellcolor[HTML]{DDDDDD}}
 \newcommand{\comment}[1]{}

\begin{document}

\title{Shoulder Surfing: From An Experimental Study to a Comparative Framework}

\author[1]{Leon Bo\v{s}njak\corref{cor1}}
\ead{leon.bosnjak@um.si}
\author[1]{Bo\v{s}tjan Brumen}
\ead{bostjan.brumen@um.si}

\cortext[cor1]{Corresponding author}
\address[1]{Faculty of Electronics and Computer Science, University of Maribor, Koro\v{s}ka cesta 46, 2000 Maribor, Slovenia}

\begin{abstract}

Shoulder surfing is an attack vector widely recognized as a real threat - enough to warrant researchers dedicating a considerable effort toward designing novel authentication methods to be shoulder surfing resistant. Despite a multitude of proposed solutions over the years, few have employed empirical evaluations and comparisons between different methods, and our understanding of the shoulder surfing phenomenon remains limited. Barring the challenges in experimental design, the reason for that can be primarily attributed to the lack of objective and comparable vulnerability measures. In this paper, we develop an ensemble of vulnerability metrics, a first endeavour toward a comprehensive assessment of a given method's susceptibility to observational attacks. In the largest on-site shoulder surfing experiment (\textit{n} = 274) to date, we verify the model on four conceptually different authentication methods in two observation scenarios. On the example of a novel hybrid authentication method based on associations, we explore the effect of input type on the adversary's effectiveness. We provide first empirical evidence that graphical passwords are easier to observe; however, that does not necessarily mean that the observed information will allow the attacker to guess the victim's password easier. An in-depth analysis of individual metrics within the clusters offers insight into many additional aspects of the shoulder surfing attack not explored before. Our comparative framework makes an advancement in evaluation of shoulder surfing and furthers our understanding of observational attacks. The results have important implications for future shoulder surfing studies and the field of Password Security as a whole.

\end{abstract}


\maketitle

\section{Introduction}

In pursuit of a solution to the age-old password security problem, a considerable effort has been made toward developing novel authentication methods aimed to replace textual passwords. Their primary challenge remains the maximization of two inherently contradicting parameters: security and usability. Papers investigating competing schemes typically encompass evaluations along these two axes, aiming to offer improvements over textual passwords in both areas simultaneously. A considerable predicament, however, is ensuring the comparability between vastly different authentication methods. To address this dilemma, studies adopted homogeneous measures, such as entropy as a measure of password strength (and by extension, the method's underlying security), or login times and recall accuracy to characterize some of the usability aspects. While none of these metrics are perfect by any means, they provide common grounds for inter-method comparison. 

Very few existing papers attempted to empirically evaluate and compare authentication methods in terms of their resistance to shoulder surfing. Instinctively, the reason could be attributed to the attack being among the less prominent ones; however anecdotal evidence suggests shoulder surfing occurs more frequently than we might think, as it can be easily perceived and carried out by an average user. Furthermore, the field of Password Security is bloated with novel authentication methods advertising to be resilient against shoulder surfing attacks, which shapes the shoulder surfing phenomenon into a well-known security problem within the human-computer interaction community.

Fair comparisons are limited due to the lack of measurements that would comprehensively and objectively characterize a given method's vulnerability to shoulder surfing attacks. Too often, simple and arbitrary measures, such as the inability to guess the password within a given number of observations, are used to determine the method "secure" against these attacks. As pointed out by \cite{Wiese15}, if all novel methods were considered secure against shoulder surfing, then comparing them along this dimension would be inconsequential. However, it is unreasonable to expect that no differences in shoulder surfing vulnerability levels would occur between any authentication methods whatsoever.

Whereas \cite{Wiese15} explores the differences in designs of shoulder surfing studies and highlights the related issues and problems, our aim is to provide a comprehensive set of metrics to allow for an objective examination of shoulder surfing resistance. In this paper, we focus on establishing a framework that can be used to compare any knowledge-based authentication methods, regardless of their underlying design. We validate our model on three vastly different authentication methods: the conventional textual passwords, a chess-based graphical authentication scheme, and a novel hybrid textual-graphical method developed for the purpose of this study. In one of the largest live-observation experiments, we perform an extensive vulnerability analysis of the considered methods, focusing on various aspects such as the input type, and the observer's intent. Beyond the first in-depth evaluation of the methods, our interpretation of the results provides valuable insight into the underlying reasons that influence a method's susceptibility to these attacks. The conclusions drawn have important implications for the design of novel authentication methods and shoulder surfing studies alike, and contribute to a better understanding of the shoulder surfing phenomena, and by extension password security as a whole.

\subsection{Literature Review}

Empirical evaluation of the shoulder surfing phenomena is rare in password research to begin with. Most often, the concerns regarding shoulder surfing attacks are addressed in papers introducing novel graphical password authentication methods. Through the evolution of authentication, resistance against shoulder surfing attacks has become an expectation, particularly for graphical passwords that were assumed to be the most susceptible to them. To this day, it is relatively common to find papers advertising a novel method as shoulder surfing resistant, though most of them do not explore that aspect beyond theoretical rationale \cite{Li15}\cite{Yakovlev}\cite{Shin}. Some of them identify the investigation of the method's resistance as a plausible direction for future research \cite{Lin}, or even presume the method is safe against the attack by the virtue of its design \cite{Khot}\cite{Rao}\cite{Sun}. This section focuses on the papers that examined shoulder surfing attacks from an empirical perspective. A detailed comparison of the selected papers can be found in Appendix A. 

\textbf{General shoulder surfing studies.} In 2009, \cite{Lashkari} conducted a survey on existing graphical authentication schemes promoted to be resistant to shoulder surfing. They described a total of sixteen papers, outlining the research problems addressed in them, as well as the methodology used, results, and potential directions for future work. Twelve papers assessed to be resistant to shoulder surfing were then selected for further comparison. While a brief overview of contemporary shoulder surfing resistant graphical authentication schemes offered some insight into shoulder surfing experiments, little effort was devoted to identifying and understanding any possible study design issues. 

The closest to exploring the challenges in design of shoulder surfing studies was the work by \cite{Wiese15}. They attempted a different approach: rather than comparing existing shoulder surfing studies in terms of results, they instead focused on \textit{how} the experiments were carried out. Through their comparison, the authors distinguished a multitude of setups and assumptions across various shoulder surfing experiments, demonstrating that simple changes in study design could have a significant impact on both validity and interpretation of the results. An analysis of different approaches also allowed them to identify several problems (such as lack of comprehensive measures) often observed in shoulder surfing research. They compiled a set of recommendations, aimed to provide future researchers a common ground to ensure comparability of their results. Advice given in this paper was taken into consideration in our study, as well.

A particularly interesting study by \cite{Kwon} examined the effect of cognitive training on the effectiveness of human adversaries. For the purpose of the study, a novel covert attentional shoulder surfing approach was designed, employing suppression of saccadic eye movement and perceptual grouping to increase adversary efficiency. A demonstration on a sample of 10 participants showed that the average shoulder surfing efficiency on a shoulder-resistant method originally proposed by \cite{Roth} increased from 44\% to 84\% in just five days. To resist covert attentional shoulder surfing, they designed an improvement of the original method, and repeated the experiment. The results showed that no participant managed to guess even 3 digits, and that all the participants missed all PIN digits in 69.8\% of the trials over the five day period. The authors concluded by warning about underestimating skilled human attackers, particularly when no countermeasures are employed to diminish the effect of sophisticated observation strategies. The same point was raised by \cite{Wiese15}.

\textbf{Shoulder surfing studies of novel and existing methods.} \cite{Jebriel} studied shoulder surfing susceptibility and usability of a recognition-based graphical password scheme using doodles as pass-images. 40 participants were divided into teams of two, consisting of victims and observers. The victims were tasked to input four pre-selected doodle passwords while the observer watched the login process. Then, the participants switched the roles and repeated the experiment. Overall, the participants managed to guess 53.6\% of all passwords; when guessing the passwords input with a mouse, they were successful in more than three out of every four cases, suggesting that the input type affects the attacker's effectiveness. 

\cite{Zakaria} proposed three shoulder surfing defence techniques for the Draw-A-Secret (DAS) recall-based graphical password scheme. In two separate experiments, susceptibility to shoulder surfing and usability were investigated. In the shoulder surfing experiment, each participant was assigned one of the four experimental groups (three defence groups and a control group), and assumed the role of an attacker trying to steal three DAS passwords (weak, medium, and strong) during individual login attempts. The results were organized into two groups based on the proportion of strokes shoulder surfed: DAS only, with Decoy Stroke defence had approximately 77\% strokes guessed, while the Disappearing Stroke and Line Snaking defences had between 40\% and 50\% strokes guessed. The authors also reported the numbers of passwords completely, and partially stolen, as well as passwords completely resistant to shoulder surfing. The effect of password strength on the guessing success was also examined. 

EvoPass, a recognition-based authentication method that evolves pass images toward shoulder surfing resistant ones, was proposed by \cite{Yu}. The authors carried out an extensive analysis of the novel scheme, including a shoulder surfing experiment, in which 20 participants observed the experimenter entering several passwords in three variations of the EvoPass system. The number of observations necessary for the participant to identify all pass images within a single attack was reported. Each participant was also asked to estimate their memory accuracy, so that the relationship between their perceived memorability and actual shoulder surfing efficiency could be explored.

A similar experiment on standard pass images was conducted by \cite{Dunphy} on mobile devices. 16 participants first enrolled in a usability study, which was followed by a shoulder surfing experiment, after they have gained some experience with the system. Analogous to \cite{Jebriel}, participants acted as both attackers and victims. The average number of observations required for a successful login was 4.5 for low- and 7.5 for high-entropy passwords. Concurrently, a model of a shoulder surfer was developed, and 10,000 attacks were simulated. The model estimated the attacker would need less than five observations to achieve one successful login; and an attacker with the means for perfect recall (e.g. camera-equipped) would be able to identify all key images 84\% of the time on average.

Another study on mobile devices examined shoulder surfing resistance of swipe passwords \cite{Cain16}. In two experiments, the participants observed video recordings of a login process. The authors revealed that the participants were significantly better at observing symmetrical patterns than asymmetrical. They also had trouble observing knight moves, suggesting that the complexity of the swipe password had impact on the shoulder surfing attack's success. Additionally, they pointed out that the guessing accuracy was much lower when there was no visual feedback (i.e. disappearing lines).

\cite{Wiese16} conducted a larger experiment to examine the shoulder surfing resistance of SwiPIN, a gesture-based swipe password method for mobile phones. Each out of 162 participants completed up to ten experimental stages in a web-based environment. In the first stage, they observed a one digit password, and in each subsequent stage they had to observe a password with one digit more than in the previous stage. To advance to the next stage, the participant had to correctly guess a random password at least three out of five times. For the first three experiments, 90\% of participants managed to observe the correct password. The success rate abruptly dropped to 56\% in experiment four, and 18\% in experiment five; only one participant managed to guess a 7-digit password three out of five times. By calculating Wilson's estimates of 95\% confidence intervals, the authors predicted which of the participants' reports were observed, and which might have been guessed. In a smaller, follow-up experiment, a similar procedure was repeated on smartphones. Through eleven stages, 19 participants were tasked to observe up to four digits, and four finger movements. A comparison of the first four experiments with the web-based study suggested that it was much harder for the participants to observe the passwords input on the phone. The authors also simulated an adversary guessing 10,000 PINs across 20 sessions in three scenarios, reporting a 90\% success within 8 to 14 observations. Finally, a simulated attack on five different schemes with a similar design was executed. The authors compared the methods in terms of success probabilities within a given number of observed sessions.

Several other studies increased the validity of their shoulder surfing experiment by performing additional evaluations. In \cite{Ho}, a proposed picture-based scheme was initially tested against a frequency of occurrence attack through a simulation. After confirming that the final "target" pictures are uniformly distributed across the locations, a standard shoulder surfing experiment was conducted on a sample of 30 participants. Despite having an unlimited number of tries when observing a recording of a login process, none of the participants managed to guess the password. \cite{Papadopoulos} developed a human visual perception algorithm to determine whether the user's keypad is visible to an observer at any given viewing position. In the shoulder surfing experiment, 21 participants worked in pairs to evaluate the estimated safety distance on a novel IllusionPIN method. Each participant was tasked to observe the login process five times from different distances. None of the attackers were able to guess the password; the authors estimated the success rate to be within the [0, 0.1329] interval. \cite{Maqsood} extensively evaluated a novel gesture-based authentication scheme called Bend passwords. They empirically compared various usability aspects of user-chosen and system-generated Bend passwords, and regular PINs. In a supplementary questionnaire, they were asked several questions pertaining to usability and security of both methods, including their perceived vulnerability against shoulder surfing attacks. In a follow-up study, 9 participants observed 8 passwords of each type in various hand position and password strength configurations, and were allowed up to three attempts to guess the password. Levenshtein distance was used to measure the similarity between the original and the guessed passwords, but no statistically significant differences were found between the methods due to a small sample size. A post-task questionnaire evaluated participants' perceived ease of shoulder surfing the passwords, and both methods were found equally difficult to shoulder surf. A variety of shoulder surfing strategies were also disclosed.

One of the largest shoulder surfing experiments to date was conducted by \cite{Aviv}. A total of 1,173 participants were recruited online (and an additional 91 locally), and were tasked to observe video recordings of victims entering 4- and 6-digit PINs and Android unlock patterns from several different angles. Additional comparisons were made with respect to screen size, hand position, and the effect of multiple over a single observation. Patterns were found to be the most vulnerable: 64.2\% uncovered the 6-point pattern within a single observation, and 79.9\% within multiple observations. In accordance with \cite{Cain16}, the figures were lower when there was no feedback (35.3\% and 52.1\% for single and multiple observations, respectively). However, 6-digit PINs appeared much harder to shoulder surf. Only 10.8\% participants guessed the PIN within one observation, and 26.5\% within multiple. Furthermore, viewing angle and phone size were shown to affect the attacker's ability to observe the password, while hand position did not. The paper complements some of the previous work done in the field, while providing insight into settings and configurations that could minimize the methods' susceptibility to shoulder surfing attacks.

Another study comparing several graphical authentication methods on mobile phones was carried out by \cite{Schaub13}. In a two-part experiment, they examined usability and susceptibility to shoulder surfing of six existing schemes, representing different types of graphical passwords. In the shoulder surfing experiment, each out of 60 participants was assigned to a graphical method group, and tasked to observe four passwords being input (strong vs. weak, live vs. video). Because of the differences between the authentication methods, the authors decided for a simple binary metric, awarding 1 to participants that guessed the correct password within three attempts, and 0 otherwise. As expected, weaker passwords were generally easier to shoulder surf than the stronger ones. The participants were significantly worse at guessing the passwords they observed on the video, likely due to fixed camera and hand positions, as well as lighting. Finally, cued-recall schemes appeared to be the most resistant, while methods containing drawing interaction (like the recall-based Pass-Go \cite{Tao}) were the most susceptible to shoulder surfing attacks.

To the best of our knowledge, only \cite{Tari} compared shoulder surfing susceptibility of a graphical authentication scheme to conventional textual passwords. Placed in the role of an attacker, 20 participants were tasked to observe four configurations of passwords (dictionary and non-dictionary textual passwords, and Passfaces input by a mouse or a keyboard). Their success rate was measured by the number of correctly guessed characters in the correct order. The results showed the participants were the least successful at guessing the keyboard-input Passfaces, while it was significantly easier for them to follow the characters input by a mouse. Surprisingly, dictionary passwords appeared more difficult to shoulder surf than the non-dictionary passwords. The authors attributed the participants' success against non-dictionary passwords to their focus on individual characters. However, a single simple measure may obstruct the real reason behind the score, as well as the actual vulnerability levels of the considered methods. In our study, a significant effort is made toward increasing the validity of obtained results by considering multiple metrics. \cite{Tari}'s main contribution, however, is the comparison of the methods' real shoulder surfing risks with the participants' perceptions. Their opinions were to some extent consistent: A post-hoc correlation analysis showed their perceptions matched the reality for mouse-input Passfaces, and dictionary passwords, while they made incorrect assumptions for the remaining two methods.

\cite{Eiband} studied the user perceptions from a different point of view. In the only real (as opposed to laboratory) shoulder surfing study, they surveyed 174 participants on their experiences related to the attack. The questionnaire inquired about the context of the attack in everyday life, type of content being observed, the perceived motivations and feelings, and reactions to the attack. Most of the questions were open-ended, allowing for a more fine-grained interpretation of the participants' diverse experiences. The analysis showed that the majority of attacks occurring in the wild were actually casual and opportunistic, with the observers rarely harboring malicious intent. Despite the attacks rarely having any serious consequences, the attackers most often observed personal data, ranging from the information about the victim's hobbies and interests, conversations and intimate details, and credentials. Consequently, the attacks elicited generally negative feelings for both parties involved, resulting in a variety of coping strategies discussed in the paper. Coupled with \cite{Muslukhov}'s finding that most malicious attacks are carried out by insiders (e.g. friends, co-workers, and family), the results provide important implications for the design of future shoulder surfing studies based on the real-world relevance of these attacks.

In a short overview of shoulder surfing experiments on existing graphical authentication methods, \cite{Cain17} were one of the few to address two key problems with shoulder surfing studies: individual scheme investigations and diverse measures. A small-scale shoulder surfing experiment comparing three existing graphical schemes provided their concept of how shoulder surfing studies should be conducted. However, three similar methods were intentionally chosen to allow for an easier comparison; no advice on how to compare vastly different methods such as textual and graphical passwords was given. Furthermore, like in all studies overviewed so far, the susceptibility to shoulder surfing was measured using simple measures, such as the proportion of the password being guessed correctly. Our work aims to expand on that.

\subsection{Main Contributions}

In this work, we make the following key contributions to the field of password security:

\begin{enumerate}
\setlength{\itemsep}{0pt}
\item \textbf{Vulnerability metrics.} To overcome challenges when evaluating shoulder surfing attacks, we establish an ensemble of vulnerability metrics. We combine individual metrics into several clusters, meant to represent different aspects of shoulder surfing attacks. The developed model is the first purposeful endeavour toward a comparable and objective measure of shoulder surfing susceptibility, laying foundations and providing common grounds for all future shoulder surfing experiments. Finally, the validity of the model is verified on four different authentication methods.

\item \textbf{Association lists.} We develop a novel textual-graphical authentication method based on associations. While a promising step forward in cognitive authentication, the method also allows us to investigate how liable personal associations are to another's guessing. Most importantly, however, with the method's hybrid design, we can observe how input type (mouse vs keyboard) influences the method's susceptibility to shoulder surfing attacks. We believe the results provide the first empirically supported evidence towards graphical passwords' predisposed vulnerability to observational attacks, as has often been pointed out in literature \cite{Aviv}\cite{Cain16}\cite{Kwon}\cite{Schaub12}.

\item \textbf{An extensive evaluation.} We perform a large-scale shoulder surfing experiment, featuring several textual and graphical authentication schemes and two types of observers (active and passive). The application of vulnerability metrics allows us an in-depth analysis of the individual factors that influence the effectiveness of these attacks. Furthermore, the normalized metrics make it easier for us to compare entirely different authentication methods, which has not been attempted before. To the best of our knowledge, this is the first shoulder surfing experiment of such scale and detail, providing crucial insight into the shoulder surfing phenomenon as a whole. 

\item \textbf{New insights.} Based on the extensive results obtained through the study, we gain a much clearer understanding of the previously only superficially examined attack on passwords. That allows us to provide empirical (rather than theoretical or even just inferential) evidence on differing shoulder surfing susceptibility levels of several authentication methods, with the employed metrics providing the tool to comprehensively rationalize the results from different viewpoints. The conclusions drawn thus have important implications not only for the evaluation of password vulnerabilities, but the field of Password Security as a whole.
  
\end{enumerate}

\section{Preliminaries}

\subsection{Threat Model}

There are various possible circumstances in which shoulder surfing attacks can be carried out. The choice of a threat model not only influences the design and evaluation of the experiment, but also has important implications for the conclusions drawn from the study. Wiese and Roth coarsely define four categories that can be drawn along two axes: (a) live versus video, and (b) single versus multiple observations \cite{Wiese16}. Our scenario assumed an opportunistic observer, such as a random adversary observing a victim's authentication process in a public space (e.g. a caf\'{e} or a library). The authors argue in favour of such weak assumptions because they model the worst-case scenario \cite{Wiese15}. If an adversary can compromise a password under such unfavorable conditions, they are more likely to be successful when their opportunities are not incidental. That can give us a broader sense of how vulnerable authentication methods really are to shoulder surfing attacks. 

Live simulations of shoulder surfing attacks were also recommended over videotaped observations \cite{Wiese15}. We highlight two arguments for our choice of live observations. First, this approach allowed us to observe the users' behaviour in the specific environment we modeled in our experiment. That way, we could gain insight into how an actual attacker might behave in the wild. More importantly, we could make an active effort to minimize the effect of variables that might have threatened the validity of the experiment. Second, adversaries are limited to their own ability to observe, encode and memorize the observed information. Since they need to share cognitive resources, their success is likely diminished when compared to a situation in which they had to focus only on one cognitive burden at a time. Such a threat model has been well described and practised in literature \cite{Jebriel}\cite{Tari}\cite{Zakaria}, and is preferred in our case for its wide applicability.

Another consideration was the role and expertise of the participant in the experiment. To model opportunistic attacks, we cast each individual participant into a role of an observer. Further, we broke the attack model down to represent two different types of opportunistic observers: deliberate and incidental; each participant was then assigned to one of the two groups. Deliberate observers were inherently malicious, and followed the authentication process with the intention to compromise the input password. Incidental observers on the other hand were emulating random passerby's that saw the password being input by chance, but did not actively try to memorize it. To that end, the former were familiarized with the authentication methods, whereas the latter would be given no prior knowledge about the schemes under study. The victim was consistently represented by one of the experimenters, who was sufficiently proficient with all considered authentication methods. With that, we aimed to provide equal conditions for all participant adversaries, while ensuring no over-estimation of security could occur. That is in line with our previous endeavour towards a broad, inclusive threat model. 

\subsection{Authentication Methods}

The main reasoning behind our choice of specific authentication methods stems from the lack of comprehensive and conclusive research on the shoulder surfing phenomenon. Proposals of novel graphical schemes in particular often address their inherent shoulder surfing vulnerability from a theoretical perspective (e.g. \cite{Li15}\cite{Shin}\cite{Yakovlev}), whilst only a few papers offer any empirical evidence to support their claims (e.g. \cite{Maqsood}\cite{Yu}\cite{Zakaria}). Furthermore, only a single study provides a direct comparison with textual passwords \cite{Tari}. We sought to close that gap by examining and comparing the susceptibility to shoulder surfing attacks of three authentication methods. A quick overview of the considered schemes is given in the next section.

\subsubsection{Textual Passwords}

For the past several decades, textual passwords have been the dominant authentication method, which can be attributed to their diverse advantages, ranging from low cost and simplicity of implementation, to convenience. Their shortcomings in security which are manifesting in a growing number of security breaches, can potentially be mitigated by users and administrators alike. As argued by Bonneau et al., no existing solution currently outperforms textual passwords in terms of security, usability, and deployability \cite{Bonneau12b}. Until a Pareto-improving authentication method is discovered and users are motivated to replace textual passwords, they are likely to remain widespread \cite{Bosnjak}.

For that reason, it is crucial to investigate all possible threat channels. To this day, the majority of password security research has been focused on password cracking. However, this approach offers a limited outlook on the entirety of the password problem, and has obscured the importance of other attacks, particularly the ones that take human factors into account. Accordingly, shoulder surfing attacks are poorly documented and studied in literature, despite textual passwords not being invulnerable to them. In our study, we conducted the first large-scale shoulder surfing experiment on textual passwords to date. Based on that, we established a benchmark for comparison and evaluation of alternative authentication methods' susceptibility to shoulder surfing attacks.

\subsubsection{Game Changer Password System}

The Game Changer Password System (GCPS) is a recent proposition by \cite{McLennan}, introducing authentication through the placement of various game pieces onto a game board in a specific order. Although the concept of using games to authenticate is not new \cite{Malempati}\cite{Tao}, the authors expand on the previous work from several viewpoints. They acknowledge the security-usability dilemma by envisioning a system that presents the user with a panel of virtual board games to choose from. The number of games available and subsequently played through can be adjusted based on the level of security needed, and the minimal usability expectations. 

\begin{figure}[h]
\includegraphics[width=.65\textwidth]{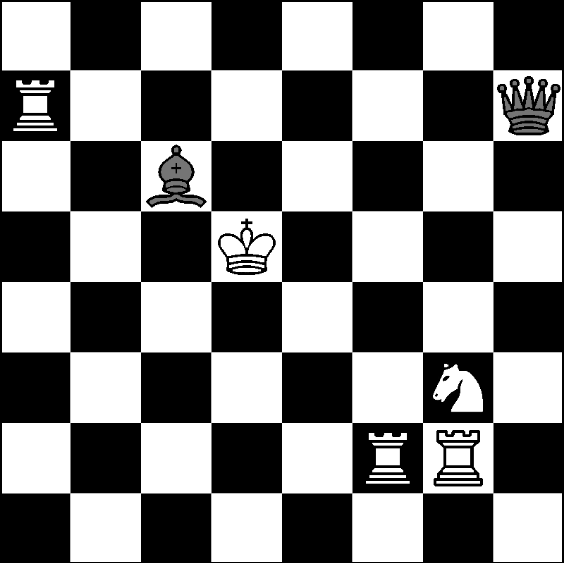}
\centering
\caption{Standard interface of the GCPS with a random, 7-character long password. A white knight is selected to be placed onto the chessboard next.}
\label{fig1}
\end{figure}

Although the GCPS works with a wide variety of board games, the authors inspected two possible implementations in their paper. We chose chess over Monopoly, namely due to its prevailing security. In chess authentication, a user is expected to put several chess pieces onto the chessboard in a predetermined order. They can choose between the standard six figures (pawn, rook, knight, bishop, queen or king) in two colors (black or white) to place them on any of the 64 positions on the chessboard. Combined, the three constructs comprise a total of 768 possible combinations per single move (as opposed to only 95 per character in textual passwords); Brumen discusses the viability of further improvements to security by expanding on the number of available constructs \cite{Brumen}. In our study, we opted for a standard chessboard with the conventional pieces as showcased in the original study by McLennan et al. \cite{McLennan}. The scheme was implemented according to their description, and with the usability aspect in mind. To make a move, a user has to select a piece from the panel of available pieces on the right side of the chessboard (the chosen piece is marked, as seen in Figure \ref{fig1}), then click on an empty field on the chessboard. The moves do not have to conform to any legal chess moves in order not to compromise security. However, while more copies of the same piece can be placed onto the game board, each square can only be occupied by a single chess piece. Consequently, all of the placed pieces are visible throughout the entire authentication process, raising concern about the method's vulnerability to shoulder surfing attacks. 

McLennan, Manning and Tuft's main contribution constitutes two experiments examining security and usability of two versions of the game-based passwords: chess and Monopoly. In Experiment 1, they compared three age groups (high school students, younger adults, and older adults) in terms of login accuracy and time on two devices. In Experiment 2, a smaller group of participants was tasked to recall and enter five different game-based passwords across 24 sessions spanning over 10 weeks. To determine whether the familiarity with the scheme affected login accuracy and reaction times, the five passwords were changed after the first 20 sessions. After both experiments, the participants also completed a questionnaire on their perceptions of the GCPS.

The authors highlight the mean login accuracy as reasonably high (77\% for Experiment 1, and 82\% for Experiment 2), despite the limited familiarity with the scheme and motivation to remember the passwords. The longitudinal study also showed that both login accuracy and reaction times gradually improved over time. When the participants were obligated to create new passwords in the middle of the study, their accuracy and performance were significantly better than the first time they entered their passwords, and subsequently improved at a quicker rate. The mean login times (28s for Experiment 1, and 11s for Experiment 2) are longer than typical users are used to, though they are comparable to the ones reported by some of the other graphical authentication schemes \cite{DeAngeli}\cite{Dhamija}\cite{Wiedenbeck}. Finally, the authors reported that the users found the GCPS to be more fun and engaging than the classic, textual passwords.

Despite the graphical schemes' well-known predisposition to shoulder surfing attacks, McLennan et al. failed to identify and address this issue as part of the possible future work \cite{McLennan}. In our study, we expand on their work by providing insight into one of GCPS' yet uninvestigated vulnerabilities. More importantly, as one of the typical and most recent representatives, the GCPS was chosen to further our understanding of the shoulder surfing phenomenon in graphical passwords altogether. 

\subsubsection{Association List Passwords}

Association Lists are a novel hybrid authentication method, designed for the purpose of this study. During authentication, the user is presented with several columns, each containing a set of numbered words. The number of columns and words in each column affects security and usability; we chose three columns of ten words in an attempt to find a reasonable middle ground. For every subsequent login, the words within each column are randomized. The user authenticates by choosing a word from each list, starting with the first and moving towards the last column. Once they have chosen a word from the last column, they can proceed with the next iteration by starting with a new word from the first column. The password can be composed of any number of words, regardless of the number of iterations and the column from which the last word was selected. The method's hybrid design allows the user to input the password using either a mouse or a keyboard. During mouse input, the user simply clicks on the words in the lists, with the interface marking the currently selected word. To input the password through a keyboard, the user has to enter the numbers pertaining to their chosen words. Because of the random word order, the input sequence is different for every login, akin to one-time passwords. The user authenticates successfully if they have selected all of the words appearing in the password in the correct order.

The main idea behind using words as the building blocks of a password is that they can represent concepts that can be easily remembered. In that regard, association lists can be considered a graphical alternative to textual passwords that were created using cognitive approaches, such as passphrases \cite{Keith}\cite{Porter}, cognitive passwords\comment{\cite{Zviran}}, associative passwords\comment{\cite{Smith}}, or the PsychoPass method \cite{Cipresso}, to name a few. Studies have shown that associative elements have positive effects on password memorability \cite{Bower}\cite{Keith}, which inspired association lists. 

\begin{figure}[h]
\includegraphics[width=.60\textwidth]{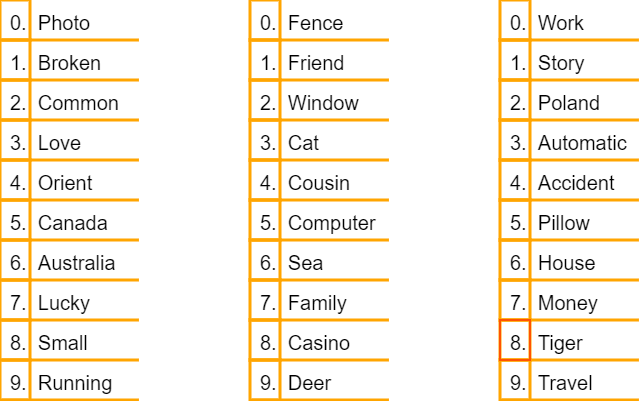}
\centering
\caption{Standard interface of the Association Lists. Only the currently selected word \texttt{tiger} is marked during mouse input. When the password is input through a keyboard, there is no visual indication of the selected words.}
\end{figure}

Generally speaking, the words in the columns could be the same for all users. However, that would not only make it difficult for some users to form associations out of words unfamiliar and unrelated to them, but it would likely make it easier for an adversary to construct a dictionary or execute a modified brute-force attack based on the concepts and word-order most likely to appear in users' passwords. Consequently, we believe each user should have a different set of words. Ideally, all words would have been pre-selected by the user prior to registration to reflect an assortment of concepts and ideas personal to them. That would allow every user to make up a story-based password built on several associations only they were familiar with. Unfortunately, like other associative approaches and even textual passwords, that makes them vulnerable to potential targeted guessing attacks \cite{Wang}.

In our case, the words in the list were pre-selected by the experimenter, who was simulating the victim in the experiment. That not only made it easier for the experimenter to memorize and enter the password during the experiment, but allowed us to showcase the use of association lists as they are intended: with the user creating and retaining their passwords based on personal associations, rather than well-established, general concepts. By extension, any potential opportunistic attackers seeing the selected words should have trouble memorizing them without having any cognitive context. 

Much like the GCPS, the association lists can be considered a graphical authentication method. Our choice to include another graphical scheme in our experiment was motivated by several factors. Instead of relying on the GCPS as the most suitable representative of graphical schemes, we aimed to increase the validity by including another scheme in the comparison. Intentionally, we chose a scheme that does not work under the same principle: while the words in association lists can act as hints (cued recall-based), the GCPS requires the user to remember the sequence of game pieces and positions without any prior clues (pure recall-based). That way, we could examine the differences in susceptibility to shoulder surfing attacks between different graphical authentication methods, as well. Finally, we took advantage of the association lists' hybrid design. By comparing the two input types, we strove to shed light on the yet unstudied nuances and aspects of graphical passwords that make them more vulnerable to shoulder surfing attacks.

\subsection{Vulnerability Metrics}

One of the significant challenges in assessing a given authentication method's susceptibility to shoulder surfing attacks is determining a set of reliable and objective metrics. In most of the studies conducted so far, shoulder surfing evaluation was not the main objective, but merely a (smaller) part of the threat analysis of an authentication scheme. As such, the authors were often interested only in whether the adversarial observer could compromise the entire password, or not \cite{Schaub13}\cite{DeLuca}. In some cases, they also measured the proportion of a password guessed correctly \cite{Kwon}\cite{Zakaria}, or the number of observations required to guess the entire password \cite{Yu}\cite{Dunphy}. The Levenshtein distance has also been used to determine the similarity between the original and the guessed passwords \cite{Maqsood}\cite{Khamis}. 

Distance metrics are often used to measure password similarity for various purposes. Most studies employ the Levenshtein distance \cite{Egelman}\cite{Schaub12}\cite{Wang}\cite{Zezschwitz}, or its variation the Damerau-Levenshtein distance \cite{Shay}, although other distance metrics, have been used as well \cite{Han}. Particularly interesting is the article by \cite{Das}, which analyzes the similarity of passwords using nine different distance metrics. The use of multiple similarity metrics additionally suggests the difficulty of selecting fair and comparable password metrics, which was also pointed out in \cite{Schaub13}. 

We believe one of the greatest challenges in designing a new similarity metric is to ensure its impartiality. That task becomes even more difficult when comparing several different authentication methods, as it was in our case. For that reason, we decided on a different approach. In this article, we aimed to establish a new model for assessing the susceptibility to shoulder surfing. Rather than implement a new similarity metric or choose one of the existing ones, we constructed an ensemble of similarity metrics to compare original passwords with the guesses, made by the observers. We classified the metrics into several clusters, based on the type of the metric. Each individual metric's score was normalized based on the original password's length to allow for a comparison between the authentication methods. The final scores for each group were calculated as the mean of all the individual scores within that group. 

It is important to consider what the scores represent. Each individual metric's value is between 0 and 1, and there were no weights placed on particular metrics when calculating the composite scores. For the majority of metrics, a higher score describes a higher similarity between the two passwords. All of the metrics for which higher scores mean a higher dissimilarity were marked with an asterisk. As our objective is to determine which authentication methods are the most resistant to shoulder surfing attacks, we are searching for scores that show the highest dissimilarity between the original and guessed passwords. 

The following similarity metrics were considered:

\subsubsection*{Password Characteristics}

\textbf{Same Chars.} The percentage of the characters in the original password appearing in the guessed password (i.e. the \% of same characters in both passwords), regardless of the order in which they occur.

\textbf{Right Spot.} The percentage of the same characters appearing in both passwords on the same positions.

\textbf{Correct First Chars.} The percentage of the original password guessed correctly from the beginning. The first wrongly guessed character in the sequence terminates the summing, even if more correct characters appear in the right positions further in the guessed password.

\textbf{Different Chars in Guess.*} The percentage of characters in the guessed password that did not appear in the original password (i.e. the \% of wrong characters in the attacker's guess).

\textbf{Longest Common Subsequence.} A subsequence is defined as any sequence derived from the password after removing some characters without changing the order of the remaining ones. The LCS is the longest subsequence, shared between the real and the guessed passwords.

\subsubsection*{Distance Metrics}

\textbf{Jaccard Index.} The number of shared characters in both passwords divided by the number of all characters appearing in either of the two passwords (i.e. intersection over union).

\textbf{Jaro-Winkler Index.} The minimum number of single-character transpositions necessary to transform the guessed password into the original password. Winkler's modification of the Jaro algorithm makes the differences between characters more significant near the start of the password string. 

\textbf{Cosine Index.} The angular distance between the two passwords, represented as points in a multidimensional space. 

\textbf{Levenshtein Index.} The minimum number of single-character edits (insertions, deletions or substitutions) required to change the guessed password into the original password. 

\textbf{N-Grams.} The extension of the Dice coefficient, which takes \textit{n}-length subsequences of characters into account. Similarity is computed as the number of shared \textit{n}-grams divided by all \textit{n}-grams appearing in both passwords. In this study, n=2 has been selected.

\subsubsection*{Guessing Order}

\textbf{Pool-based.*} The attacker assumes that they guessed the correct characters (but not necessarily their position) within the password. Initially, they would attempt all possible permutations of the guessed characters. If the original password is not found, they proceed by gradually substituting characters of the guessed password and subsequently trying all permutations.

\textbf{Position-based.*} In this case, the attacker believes they have seen correct characters appearing at correct positions in the password. If a guessed character does not appear at the guessed position in the original password, they substitute it with different characters until they find a match. Similar to the pool-based guessing order, the substitutions are gradual, starting with replacing only one password character, and later on more characters at a time until the original password is found. 

\textbf{Entropy.*} The decrease in password unpredictability from the original password to the guessed password. A given password's entropy can be calculated as \(log_2 (p^l)\), where \textit{p} is the size of the character pool, and \textit{l} is the length of the password. Previous studies have criticized entropy as a measure of password strength, demonstrating that it can over- or understate the effective security of a password \cite{Bonneau12a}\cite{Weir}. However, improvements are not applicable in our case, as they would have required existing corpora of passwords \cite{Kelley} or dictionaries \cite{Ma}, which we cannot currently provide for chess- and association-based passwords. For that reason, a suboptimal measure was chosen to provide some general sense of password strength.

\medskip

When comparing individual characters, the similarity metrics do not take the degree of correctness into account. For example, if an observer states that the lowercase letter 'a' appears in the password, while the original password contains the uppercase letter 'A', the metrics would pronounce their guess incorrect. They did not consider the fact that the observer has guessed the correct key, but did not apply the correct modifier (Shift or Caps Lock in this case). This simplification is even more pronounced for the GCPS with a character pool size of 768, and each character comprised of a figure, its color, and position on the chessboard. If the observer correctly guesses the figure and its position, but misses the color, the result of the metrics will be the same as if he had incorrectly guessed all three. 

We considered partially correct guesses by introducing an adaptation to password characteristics and distance metrics. Character pools were divided into several disjoint groups. For textual passwords, we identified two groups: the pressed key (size of 49), and the modifier (size of 3, for no modifier, Shift, and Alt Gr). The GCPS passwords were broken into three groups: the figure (size of 6), the color (size of 2), and the position (size of 64). Association list passwords could not be further subdivided into smaller groups, because the words in the pool have no common points. We calculate the characteristics and distance metrics separately for every character pool group, then apply weights to each group depending on its size. The normalized sum of the partial similarity metrics becomes the final score of the adjusted similarity metric for a given authentication method.

\section{Research Methodology}

\subsection{Research Questions}

In this study, we endeavoured to expand on the currently much unexplored domain of shoulder surfing. We were particularly interested in how susceptibility to these attacks could be measured objectively, and applied the developed metrics on a range of textual and graphical authentication methods. The main aim of the study was to provide a fair comparison of textual and graphical schemes from the perspective of shoulder surfing, while also considering the input method. For that reason, we chose the hybrid scheme association lists to observe the differences in susceptibility to observation attacks when the same passwords are entered using either a mouse\textsubscript{(M)}, or a keyboard\textsubscript{(K)}. In this regard, our work can be considered a continuation of the previous research efforts by \cite{Tari}. 

For each considered authentication method (textual passwords\textsubscript{(K)}, GCPS\textsubscript{(M)}, and association lists\textsubscript{(K,M)}), we executed the experiment in two configurations: with the participants as active, and as passive observers. After this classification we can explain the objectives of our study as follows: 

\begin{enumerate}
\setlength{\itemsep}{0pt}
\item Establishing the actual shoulder surfing vulnerability levels of textual passwords, the GCPS, and two configurations of association list passwords (keyboard and mouse input), using the newly established vulnerability metrics,

\item Determining the significant differences in the vulnerability levels between all considered authentication methods,

\item Evaluating the effect of password input (keyboard and mouse input) on the shoulder surfing vulnerability, particularly on the example of association list passwords, and

\item Comparing the efficiency of shoulder surfing attacks between two types of observers (active and passive).
  
\end{enumerate}

\subsection{Participants}

The study collected 274 valid responses (193 Male, 81 Female). Participants varied from the age of 18 to the age of 25, with a mean age of 20.5 and were all undergraduate students in Computer Science, Media Communication, and Electrical Engineering. All participants had (corrected-to-)normal eyesight, and reported they spent on average 4.91 hours a day on the computer (SD = 2.77h). The majority of participants had a computer science background, used password-based systems on a daily basis and were familiar with authentication in general. These attributes boosted their credibility as potential shoulder surfers, which is compliant with our threat model. Whereas video game players enjoying fast-paced games would qualify even better for the role of shoulder surfers \cite{Kwon}, we decided against such a sample because we were interested in the differences between active and passive observers. For a scenario in which half of the sample remains uninformed about their role in the experiment, video gamers are not suitable. Instead, a more general, yet more proficient than average sample was deemed most appropriate.

\subsection{Experimental Procedure}

The experiment was conducted in a controlled laboratory environment to avoid any possible distractions. We chose the between-subject design, meaning that each participant was assigned to one of the eight experimental groups. Each group had at least 30 participants. 

\textbf{Before the experiment:} A short pilot study with 12 participants was carried out prior to the execution of the experiment. We assigned two participants per experimental group (initially, we did not distinguish between different types of input, so association lists comprised a single group), then executed the experiment according to the original experimental design. Once the pilot experiment was completed, we gathered all participants, and educated them about the purpose of the study, as well as the reasons behind some of the experiment design choices. Then, participants were asked a set of pre-prepared questions about the experiment. These questions mainly concerned the experimental setup, the participants' opinion and impression of the study, and provided the chance for them to give additional suggestions. We concluded the pilot study with an open discussion, during which we documented all key points raised.

After the pilot study, participants' responses and suggestions were reviewed. Based on the results, we made several changes to the original experiment design. Most notably, we split the association list group into two separate groups, based on input type. Despite both experimental groups representing the same authentication method, the input types actually exhibit two different authentication behaviors. By making this change, we could directly observe the effect of input type on the susceptibility to shoulder surfing attacks. Next, we reviewed the information participants of each group would learn prior to the execution of the experiment. We also moved the \textit{password recall stage} immediately after the \textit{shoulder surfing stage} rather than the \textit{data collection stage}, to decrease potential memory decay. By eliminating such cognitive noise, the observed password should remain in short-term memory, which is compliant with our threat model. Finally, a couple of minor adjustments were made to the exit questionnaire: input field validation was implemented, and a field for number of hours spent daily on the computer was added.

All participants were notified about the time and place of the experiment at least a week prior to its execution. On the day of the experiment, the laboratory environment was prepared, and the experimental equipment was tested. The briefing room was equipped with a projector and a screen for the introductory demonstration. A smaller experiment room was connected directly to the briefing room, and contained a computer equipped with an Intel i5-8500 (@4GHz), 8GB of RAM (DDR4 2400), a Win 10 Pro (x64) OS, and running the experiment on the PHP platform (v7.1.14), and a standard 17" LCD computer screen with a 1366x768px resolution. Half a meter behind and on each side of the experimenter, two spots were marked on the floor for the participants to stand on during the experiment to ensure the same viewing angles as well as hand and mouse positions for all participants.

\textbf{During the experiment:} The experiment was executed for each experimental group separately. We organized the experimental procedure into four stages:

\textit{Briefing stage.} Depending on the experimental group, the participants were briefed about the experiment to varying levels of detail. The members of the active experimental groups were introduced to the authentication scheme through a short demonstration. They were informed about the purpose of the study, and were asked to put themselves in the role of an attacker who wants to compromise the password. They were also shown several possible attack strategies. Contrarily, the participants in the passive groups were only instructed to carefully observe the experimenter and what they are doing on the computer. We did not reveal the nature of the study to the passive observers, and we were cautious not to mention any of the authentication methods throughout the briefing. All participants were asked to turn off their mobile devices, and wait in the briefing room until they were called. One of the experimenters stayed in the waiting room to ensure none of the participants communicated with anyone outside the room. 

\textit{Shoulder surfing stage.} The participants entered the experiment room two by two directly from the briefing room. They stepped on the designated spots behind the experimenter, and assumed a comfortable position from which they could see the screen and the keyboard clearly. They were not allowed to lean forwards, and the experiment did not start until both participants proclaimed they were ready to begin. Depending on the experimental group that the participants were members of, the experimenter then completed the authentication process using one of the four considered authentication methods. Throughout the process, the experimenter navigated to the login site, entered the username, followed by the password, and clicked on the login button that displayed a welcome message upon entering the correct credentials. The experimenter was careful to input the credentials at a moderate speed, imitating the average user.

The entered passwords were the same for all participants shoulder surfing a given authentication method. Considering the character pool sizes differed between the three schemes (95 for textual passwords, 768 for the GCPS, and 10 for association lists), we had to select passwords of different lengths (11 for textual passwords, 7 for the GCPS, and 21 for association lists) to ensure the methods were comparable in terms of security (i.e. number of all possible combinations). Password lengths were not chosen arbitrarily, but aimed to provide a sufficient level of security against exhaustive attacks. To determine the appropriate threshold, we considered the password cracking speeds achievable by contemporary computers. In a recent study, \cite{Li18} reported MD5 hash cracking speeds of up to 16 GH/s using their high-order reconfigurable processing unit. A benchmark test employing a rig of 8 nVidia GTX 1080 Ti GPUs achieved over 307 GH/s for MD5 hashed passwords \cite{Gosney}. Taking the current cracking speeds into account, we estimated $\sim10^{21}$ combinations to be a search space of a sufficient size to withstand modern day cracking techniques.

\textit{Password recall stage.} Following the shoulder surfing, the participants were then asked to recall the passwords that they have seen. Each participant was given access to a computer, on which they could enter the memorized password. They were encouraged to input password characters even if they were not fully certain whether they appeared in the password. We measured the login time, starting with the participant opening the login page, and ending with the participant clicking on the login button. The time measures the duration of the entire authentication process, including the recall; it would have also been sensible to measure only the password entry time.

\textit{Data collection stage.} Participant data was collected through the means of an exit questionnaire. The responses were recorded (along with the recalled passwords) for later analysis. During the last two stages, the experimenter was available to aid with any potential technical difficulties, or reiterate the instructions.

\textbf{After the experiment:} The participants did not return to the briefing room, but instead used another exit to prevent interaction with the participants that have not yet undergone the experimental procedure.

\subsection{Statistical Analysis}

After collecting the data, we calculated vulnerability scores from the participants' guesses. We analyzed the obtained data, evaluating the login times, shoulder surfing vulnerability scores, and the questionnaire responses. When comparing more than two groups, we used the Kruskal-Wallis H test to determine whether there are any statistically significant differences between the groups. We chose this non-parametric test because the data was not normally distributed for most of the considered methods, and met other assumptions required (authentication methods are categorical, independent groups, vulnerability metrics are measured on a continuous scale, and no participants were included in the experiment more than once). To compare specific pairs, we used a post hoc test, namely the Bonferroni-adjusted Mann-Whitney U test. Finally, to measure the magnitude of the observed differences, we calculated the effect sizes for Mann-Whitney tests, using Cohen's criteria to interpret the results.

\section{Results}

The result section was organized into several segments, aimed to report the findings and address the research questions systematically. Initially, we focused on evaluating and comparing the four considered authentication methods (textual passwords, the GCPS, and the association lists with either keyboard or mouse input). Then, we split the data based on the type of the observer, and compared the methods separately for active and passive observers. Finally, we were interested in comparing the active and passive observers for each authentication method. In our comparisons, we remark on login times, however, the bulk of our analysis remains focused on vulnerability scores. In this regard, we primarily examine the composite vulnerability metrics, which provide a broad sense of the methods' susceptibility to shoulder surfing attacks. Additionally, we highlight any particularly interesting results of the individual vulnerability metrics. Notwithstanding, means and standard deviations of individual and composite metrics for all authentication methods are reported in Table \ref{table:1}. 

  {\setlength\doublerulesep{0.4pt}

  \begin{table*}[t]
  \begin{adjustwidth}{-1.5in}{-.5in}  
  \centering
  \begin{tabular}{ clccccccccc } 
  \toprule[1pt]\midrule[0.3pt]
    & & \multicolumn{2}{c}{\textbf{Textual}} & \multicolumn{2}{c}{\textbf{GCPS}} & \multicolumn{2}{c}{\textbf{List\textsubscript{(keyboard)}}} & \multicolumn{2}{c}{\textbf{List\textsubscript{(mouse)}}}  \\\midrule
    \# & \textbf{Metric} & Active & Passive & Active & Passive & Active & Passive & Active & Passive \\ \midrule[1pt]
    
    L1 & Length Dif* & .1788 {\tiny (.13)} & .3366 {\tiny (.19)}& .1384 {\tiny (.12)}& .1627 {\tiny (.15)}& .\textbf{4223} {\tiny (.21)} & \textbf{.4224} {\tiny (.20)}& .4080 {\tiny (.18)}& .3911 {\tiny (.24)} \\\midrule
    C1 & Same Chars & .4538 {\tiny (.10)} & \textbf{.2840} {\tiny (.14)} & .4423 {\tiny (.21)} & .3763 {\tiny (.19)} & \textbf{.4010} {\tiny (.15)} & .3441 {\tiny (.13)} & .4530 {\tiny (.13)} & .4300 {\tiny (.19)}  \\
    C2 & Correct First & .1452 {\tiny (.14)} & .0412 {\tiny (.09)}& .1682 {\tiny (.18)}& .1164 {\tiny (.13)}& \textbf{.0451} {\tiny (.06)} & \textbf{.0061} {\tiny (.02)}& .1145 {\tiny (.07)}& .0216 {\tiny (.04)} \\
    C3 & Right Spot & .2110 {\tiny (.12)} & .1053 {\tiny (.10)} & .2423 {\tiny (.21)} & .1852 {\tiny (.16)} & \textbf{.0714} {\tiny (.06)} & \textbf{.0230} {\tiny (.03)} & .1338 {\tiny (.06)} & .0404 {\tiny (.04)} \\
    C4 & LCS & .3943 {\tiny (.09)} & .2351 {\tiny (.11)} & .3678 {\tiny (.17)} & .3251 {\tiny (.18)} & \textbf{.2393} {\tiny (.09)} & \textbf{.1982} {\tiny (.07)} & .3192 {\tiny (.08)} & .2136 {\tiny (.09)}\\
    C5 & Dif in Guess* & .4637 {\tiny (.17)} & \textbf{.5795} {\tiny (.19)} & \textbf{.4959} {\tiny (.24)} & .5679 {\tiny (.19)} & .2979 {\tiny (.16)} & .4051 {\tiny (.13)} & .2252 {\tiny (.15)} & .3241 {\tiny (.14)} \\
    \textbf{C} & \textbf{Characteristics} & .3481 {\tiny (.10)} & \textbf{.2172} {\tiny (.10)} & .3450 {\tiny (.18)} & .2870 {\tiny (.16)} & \textbf{.2918} {\tiny (.06)} & .2333 {\tiny (.06)} & .3591 {\tiny (.06)} & .2763 {\tiny (.07)}\\\midrule
    
    D1 & Jaccard & \textbf{.3520} {\tiny (.10)} & \textbf{.2199} {\tiny (.12)} & .3600 {\tiny (.20)} & .2949 {\tiny (.17)} & .3555 {\tiny (.13)} & .3028 {\tiny (.12)} & .4142 {\tiny (.12)} & .3677 {\tiny (.15)}\\
    D2 & Jaro-Winkler & .6379 {\tiny (.13)} & \textbf{.5024} {\tiny (.13)} & \textbf{.5705} {\tiny (.23)} & .5286 {\tiny (.22)} & .6221 {\tiny (.18)} & .5435 {\tiny (.14)} & .7814 {\tiny (.13)} & .5715 {\tiny (.19)}\\
    D3 & Cosine & .5558 {\tiny (.13)} & \textbf{.3651} {\tiny (.17)} & \textbf{.4798} {\tiny (.22)} & .4102 {\tiny (.19)} & .5303 {\tiny (.13)} & .4536 {\tiny (.13)} & .5965 {\tiny (.11)} & .5278 {\tiny (.13)}\\
    D4 & Levenshtein & .2983 {\tiny (.13)} & .1806 {\tiny (.10)} & .2907 {\tiny (.20)} & .2228 {\tiny (.16)} & \textbf{.1926} {\tiny (.09)} & \textbf{.1457} {\tiny (.06)} & .2846 {\tiny (.10)} & .1567 {\tiny (.08)}\\
    D5 & N-grams & .3029 {\tiny (.13)} & .1771 {\tiny (.10)} & .3175 {\tiny (.21)} & .2432 {\tiny (.17)} & \textbf{.2176} {\tiny (.15)} & \textbf{.1319} {\tiny (.06)} & .2833 {\tiny (.09)} & .1841 {\tiny (.17)}\\
    \textbf{D} & \textbf{Distance} & .4294 {\tiny (.11)} & \textbf{.2890} {\tiny (.11)} & .4037 {\tiny (.19)} & .3399 {\tiny (.17)} & \textbf{.3836} {\tiny (.10)} & .3155 {\tiny (.08)} & .4720 {\tiny (.09)} & .3616 {\tiny (.09)}\\\midrule
    
    G1 & Pool Guess* & .9746 {\tiny (.11)} & .9760 {\tiny (.10)} & .9335 {\tiny (.14)} & .9273 {\tiny (.13)} & \textbf{.9824} {\tiny (.08)} & .9691 {\tiny (.01)} & .9673 {\tiny (.06)} & \textbf{.9927} {\tiny (.09)}\\
    G2 & Position Guess* & \textbf{.9716} {\tiny (.14)} & .9844 {\tiny (.10)} & .9090 {\tiny (.17)} & .9167 {\tiny (.13)} & .9562 {\tiny (.09)} & .9649 {\tiny (.02)} & .9189 {\tiny (.07)} & \textbf{.9883} {\tiny (.09)}\\
    G3 & Entropy* & .2111 {\tiny (.15)} & \textbf{.4369} {\tiny (.15)} & .1384 {\tiny (.12)} & .1627 {\tiny (.15)} & \textbf{.4223} {\tiny (.21)} & .4224 {\tiny (.20)} & .4080 {\tiny (.18)} & .3911 {\tiny (.24)}\\
    \textbf{G} & \textbf{Guessing Order*} & .7191 {\tiny (.10)} & \textbf{.7991} {\tiny (.09)} & .6603 {\tiny (.12)} & .6689 {\tiny (.11)} & \textbf{.7869} {\tiny (.09)} & .7855 {\tiny (.07)} & .7647 {\tiny (.07)} & .7907 {\tiny (.09)}
    
     \\\midrule[0.3pt]\bottomrule[1pt]
  \end{tabular}
  \end{adjustwidth}
  \caption{Individual and composite vulnerability metrics' means and standard deviations for all authentication methods and both observer types. The best active and passive scores are bolded for each metric. Metrics marked with an asterisk are complementary.}
  \label{table:1}
  \end{table*}
   }

\subsection{Comparison of authentication methods}

\subsubsection{Login Times} 
Pairwise comparisons showed that there were significant differences between all authentication methods, except the two variations of the association lists (\textit{U} = 2,609, \textit{p} = .414). As expected, participants in the textual password group were decisively faster (Mdn = 21.92s) than the others. That can be partially attributed to their familiarity with the scheme. However, it also showcases textual passwords' ease of use, particularly when compared to graphical alternatives. For example, participants in the GCPS group took almost thrice as long (Mdn = 61.58s), despite chess passwords being several characters shorter. Participants shoulder surfing association list passwords took even longer, with the group inputting the passwords through the keyboard (Mdn = 90.15s) being only slightly faster than the group using the mouse (Mdn = 96.09s).

\begin{figure}[h]
\includegraphics[width=.6\textwidth]{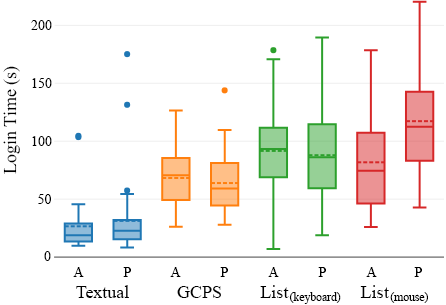}
\centering
\caption{The distribution of login times by authentication method and observer type (A - Active, P - Passive).}
\end{figure}

Based on that, we can conclude that the input type does have some effect on the input speed, though not as significant as we would have initially expected. Instead, usability factors are much more dependent on the way the information the user has to memorize is encoded, and their ability to recall and subsequently input it quickly and efficiently. Textual passwords offer the advantage because mapping the recalled concept into the password construct is straightforward: inputting a textual password's character is as simple as pressing a single key on the keyboard. On the other hand, inputting a single GCPS character is considerably slower, because the user has to select the correct character and move it onto the correct position on the chessboard. Word selection in randomized association lists likely takes even longer, considering the user has to first find the correct word before they can select it. Furthermore, list passwords were comprised of more characters because of the method's small character pool.

Another important factor is the cognitive burden associated with password recall. Unfortunately, by measuring the time passed from the start to the end of the authentication process, we were unable to determine the amount of time the users dedicated to recalling the password they have observed, as opposed to entering it. Perhaps by recording the password input time as well, we could have deduced which method's observed passwords were the easiest for the participants to recall. This remains a subject for future studies, as well as a valid point to consider in any subsequent shoulder surfing experiments.

\begin{figure}
  \begin{adjustwidth}{-1.25in}{-.5in}  
\begin{minipage}[t]{.6\textwidth}
    \centering
    \subfloat[Password Characteristics]{{\includegraphics[width=.9\linewidth]{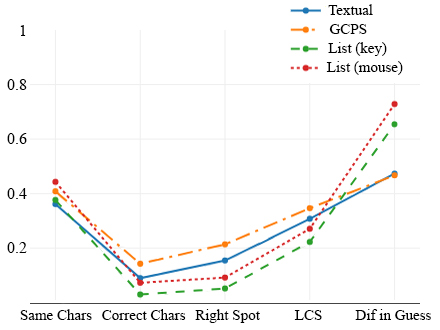}}}
    \subfloat[Distance Metrics]{{\includegraphics[width=.9\linewidth]{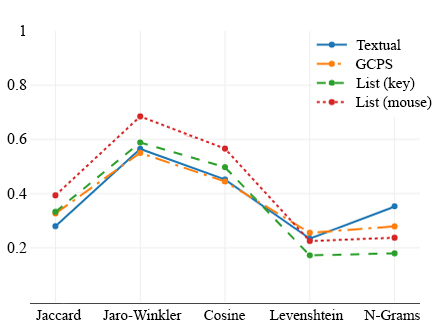}}}
    \subfloat[Guessing Order]{{\includegraphics[width=.675\linewidth]{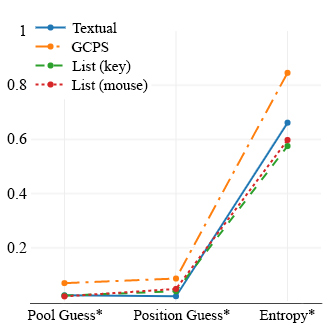}}}
\end{minipage}
\end{adjustwidth}
  \caption{Comparison of metric scores across all authentication methods. Metrics marked with an asterisk are complementary. Consequently, lower scores for all metrics denote the method more resistant to shoulder surfing attacks.}
  \label{fig:2}
\end{figure}

\subsubsection{Vulnerability Metrics}

\textbf{Password Characteristics.} This set of metrics aimed to encompass several defining features of an attacker's guessing tactics. In general, it is meant to give us a broader idea of how effective the majority of the guessers' approaches were against specific methods; a drill-down of individual metrics provides insight into how exactly the participants approached the guessing task, and how successful their strategies were.

One of our key findings is that while textual methods generally appear less vulnerable to shoulder surfing attacks than their graphical counterparts, these differences are often too small to be considered significant. Most notably, no differences in shoulder surfing susceptibility have been reported between textual passwords and the GCPS (\textit{U} = 2,490, \textit{p} = 1), the prime representatives of the two opposing knowledge-based authentication method types. These results may initially suggest that the graphical aspect might not be the decisive factor influencing the method's vulnerability to shoulder surfing attacks. However, the comparisons with association lists showed that the mouse-input list passwords performed significantly worse than their textual counterparts (\textit{U} = 3,440.5, \textit{p} \textless .001), and the traditional textual passwords (\textit{U} = 2,958, \textit{p} = .048). The difference between the two variations of the same authentication scheme in particular showcases that the graphical component does have a negative effect on the method's vulnerability to shoulder surfing attacks. However, other security (e.g. chess' large character pool) and usability (e.g. lists' random word order) factors may diminish this effect to some degree. 

\textit{Same Chars.} When considering the proportion of correctly guessed characters in the passwords, we observed practically no differences between the methods. The participants guessed more characters for the methods with smaller character pools, such as association lists; however, they made a considerable greater proportion of partially correct guesses for larger-pool methods, such as the GCPS. In other words, the attackers managed to construct a proportionally similarly-sized pool of characters they considered as possible candidates to appear within the targeted password. This might suggest that graphical passwords are not necessarily that much easier to shoulder surf, as it has been repeatedly pointed out in literature. The differences likely get smaller as the passwords become more complex. Our choice of strong passwords was influenced by our motivation to study the shoulder surfing vulnerability on passwords resistant to contemporary brute-force attacks. In this case, the visual aspect of graphical passwords did not significantly contribute to the attackers seeing \textit{and} remembering the observed information. The cognitive load was, for all considered methods, too high for the attackers to shoulder surf and memorize more than just a few characters.

  {\setlength\doublerulesep{0.4pt}
  \begin{table*}[t]
  \begin{adjustwidth}{-.8in}{-.5in}  
    \centering
  \begin{tabular}{ clcccccc|ccccc } 
  \toprule[1pt]\midrule[0.3pt]
    & & \multicolumn{6}{c}{\textbf{Authentication Method}} & \multicolumn{4}{c}{\textbf{Active vs Passive}} \\\midrule
    \# & \textbf{Metric} & \textbf{P-C} & \textbf{P-L\textsubscript{k}} & \textbf{P-L\textsubscript{m}} & \textbf{C-L\textsubscript{k}} & \textbf{C-L\textsubscript{m}} & \textbf{L\textsubscript{k}-L\textsubscript{m}} & \textbf{P} & \textbf{C} & \textbf{L\textsubscript{k}} & \textbf{L\textsubscript{m}} \\ \midrule[1pt]
    
    L1 & Length Dif & \sig .006 & \sig 0 & \sig 0 & \sig 0 & \sig 0 & 1 & \sig 0 & .534 & .73 & .588 \\\midrule
    C1 & Same Chars & 1 & 1 & \sig .024 & 1 & .834 & .084 & \sig 0 & .079 & .109 & .623 \\
    C2 & Correct First & \sig .03 & \sig .006 & 1 & \sig 0 & \sig .024 & \sig 0 & \sig 0 & .231 & \sig .001 & \sig 0  \\
    C3 & Right Spot & .792 & \sig 0 & \sig .024 & \sig 0 & \sig 0 & \sig 0 & \sig 0 & .245 & \sig 0 & \sig 0 \\
    C4 & LCS & 1 & \sig 0 & .318 & \sig 0 & .078 & \sig .018 & \sig 0 & .117 & \sig .039 & \sig 0 \\
    C5 & Dif in Guess* & 1 & \sig 0 & \sig 0 & \sig 0 & \sig 0 & \sig .036 & \sig .025 & .263 & \sig .008 & \sig .011 \\
    \textbf{C} & \textbf{Characteristics} & 1 & 1 & \sig .048 & 1 & .972 & \sig 0 & \sig 0 & .154 & \sig 0 & \sig 0 \\\midrule
    
    D1 & Jaccard & 1 & .102 & \sig 0 & 1 & \sig .006 & .054 & \sig 0 & .152 & .11 & .2 \\
    D2 & Jaro-Winkler & 1 & .96 & \sig 0 & 1 & \sig 0 & \sig 0 & \sig 0 & .32 & \sig .009 & \sig 0 \\
    D3 & Cosine & 1 & .96 & \sig 0 & .096 & \sig 0 & \sig .018 & \sig 0 & .193 & \sig .025 & \sig .029 \\
    D4 & Levenshtein & 1 & \sig .042 & 1 & \sig .024 & 1 & \sig .018 & \sig 0 & .163 & \sig .012 & \sig 0 \\
    D5 & N-grams & 1 & .072 & 1 & \sig .006 & 1 & \sig .006 & \sig 0 & .156 & \sig 0 & \sig 0 \\
    \textbf{D} & \textbf{Distance} & 1 & 1 & \sig .006 & 1 & .084 & \sig 0 & \sig 0 & .137 & \sig .002 & \sig 0 \\\midrule
    
    G1 & Pool Guess* & 1 & 1 & 1 & \sig .036 & \sig .018 & .486 & .084 & .64 & .554 & .054 \\
    G2 & Position Guess* & \sig 0 & .12 & \sig 0 & .066 & 1 & \sig .024 & .84 & .857 & \sig .005 & \sig 0 \\
    G3 & Entropy* & \sig 0 & .114 & .438 & \sig 0 & \sig 0 & 1 & \sig 0 & .534 & .73 & .588 \\
    \textbf{G} & \textbf{Guessing Order*} & \sig 0 & .45 & 1 & \sig 0 & \sig 0 & 1 & \sig 0 & .457 & .847 & .247 \\\midrule[0.3pt]\bottomrule[1pt]
    
  \end{tabular}
  \caption{Bonferroni-corrected \textit{p}-values for pairwise comparisons between all authentication methods, and for pairwise comparisons between the active and passive observer groups within each considered method. All \textit{p}-values, significant at .05 level, are shaded. Metrics marked with an asterisk are complementary.}
  \label{table:2}
  \end{adjustwidth}
  \end{table*}
   }
   
\textit{Correct First.} There were, however, significant differences in the number of correctly guessed characters, starting from the beginning of the password. Despite association lists having the smallest pool, the participants actually guessed the smallest amount of first characters. On the other hand, although the GCPS has the largest pool of the three considered methods, the participants guessed significantly more first characters than even the participants of the textual password group. We observed varying levels of guessing difficulty between the authentication methods, owing to their design and usability aspects. For example, the reason why few first characters were guessed correctly for association lists, is because the participants struggled with the changing word order: the word \texttt{deer} appearing in the first position during the shoulder surfing stage likely didn't reappear in the same position during the password recall stage. Furthermore, the keyboard group performed even worse because of the additional cognitive burden of having to map the observed strokes to the corresponding words. As a result, many participants might have tried to simply blind-guess the list passwords. Contrarily, the GCPS group was much more accurate. Aside from the graphical representation allowing them to see the password easier, we believe the participants could attempt to remember the initial sequence of characters by imagining them as the moves on the chessboard. A similar approach could not easily be done with association lists - the attacker would not have insight into personal associations constructed by the victim.

\textit{Right Spot.} Measuring the proportion of correct characters appearing in correct positions within the passwords yielded similar results. Even though the participants could "guess" the correct characters appearing in the association list passwords (due to the small character pool), very few participants actually managed to place them in the right positions. That reinforces our assumption that the participants were unable to guess the associations in the victim's mind, thanks to every set of words being personal to the victim. Conversely, they were significantly better at guessing where the correct characters appeared in textual and chess passwords, most likely because they could to some degree predict the following characters based on the previous ones. Much like the number of correct first characters, this measure suggests the participants are more likely to blind-guess characters when they are not certain in their own guess. Moreover, these two metrics shed light on the reason \textit{why} the differences in susceptibility between the methods are fairly small, as the metric measuring the number of same characters implies. While the type of a method (textual or graphical) influences the susceptibility to shoulder surfing attacks to some degree, other features of the method also matter. That is why attackers can discern a similar amount of information when shoulder surfing a highly visual but mathematically robust method such as the GCPS, or the much less intuitive for the attacker, yet easier to blind-guess (due to a small pool of character) method, such as the association lists.

\textit{LCS.} An observer's degree of certainty in their guess can (to some extent) be described by the longest common substring. The main idea is that while following the password being input, the attacker might miss some of the characters. The more characters they miss, the greater the chance they might not have actually seen and memorized the characters, but were instead blind-guessing them. Comparison of authentication methods revealed it was considerably harder for the participants of the list-key group to follow the sequence of correct characters, than for the other three groups. Such results were to be expected; seeing the correct association list characters being input through the keyboard is particularly difficult, because the attacker needs to observe the number on the keyboard, and then map it to the correct word on the screen. Consequentially, we believe most list-key guesses made were actually blind. For all the other three methods, the participants displayed a significantly higher degree of certainty in their guesses. The visual aspect of the two considered graphical passwords likely contributed to the increased level of certainty. The reason why the participants could follow sequences of textual characters with equal ease might be due to them trying to guess the next sequence of characters. Doing the same with list-key would in fact be counterproductive because of association interference.

\textit{Dif in Guess*.} Finally, we considered the number of wrong characters in the password, penalizing the participants who were making too many blind guesses. As expected, association lists scored the worst, mainly because the attackers have the highest chance of guessing a correct character, even if they were making a blind guess. Surprisingly, despite vastly different character pools, the participants guessed only marginally less wrong characters for textual passwords than for chess passwords (\textit{U} = 2,397, \textit{p} = 1). This could suggest that the participants' guesses are generally more accurate for graphical passwords, which is further supported by the fact that they made more wrong guesses for the list-key than the list-mouse (\textit{U} = 1,770, \textit{p} = .036), despite both representing the same authentication method and only differing in input type.

\textbf{Distance Metrics.} These metrics attempt to measure the distance between two target strings, which in our case represent the victim's original password and the attacker's guessed password. A collection of most often used distance metrics in literature is used to provide a comprehensive perspective on the effectiveness of participants' guesses. The results of individual metrics are shortly described to characterize the differences in scores, depending on each metric's focus.

Consistently with the characteristics metrics, this cluster of measures also indicates no significant differences between the methods, particularly between the textual passwords and the graphical GCPS (\textit{U} = 2,401, \textit{p} = 1), where such disparities were expected to appear. Much like for the previous cluster, the only notable differences were between the list-mouse and the two non-graphical authentication methods. Once again, the list-mouse appeared more susceptible to shoulder surfing attacks than the list-key (\textit{U} = 3,321.5, \textit{p} = 0), and the textual passwords (\textit{U} = 3,121, \textit{p} = .006). The consistency of the results between the two clusters suggests either of the two (or both) sets of metrics can be used to evaluate a method's susceptibility to shoulder surfing attacks. Overall, the results captured by the distance metrics indicated less discrepancies between the methods, and were more homogeneous between the individual distance metrics.

\textit{Jaccard.} Not surprisingly, we found practically no differences between the authentication methods when measuring the distance between the two passwords using the Jaccard index. A rather simple measure, this index heavily relies on the size of the character pool, which in our case means that methods such as the GCPS would have a clear advantage. However, after adjusting the metrics to count partial guesses as well, the differences were severely diminished. That goes on to support our previous findings that the methods with larger character pools are not necessarily more resistant against shoulder surfing attacks simply by virtue of their character pool size. Instead, graphical passwords were again shown to be slightly more susceptible to these attacks, even if negligibly so, barring the significantly worse list-mouse method.

\textit{Cosine \& Jaro-Winkler.} Both cosine and Jaro-Winkler indexes are in agreement with the previous metric. The Jaro-Winkler metric was particularly interesting because it only takes correct guesses into account if they are within a given distance. In our case, the attackers were equally successful at placing their guessed characters in the vicinity of the correct characters within the password, for all methods. However, considering the variable character pool sizes, there is a greater chance that more of these guesses were blind for association lists than the other two methods. In conjunction with accidental correct guesses, we believe the graphical component of the list-mouse method allowed for a further increase in the participant's guessing effectiveness.

\textit{Levenshtein.} Levenshtein index is focused on the number of edits necessary to change the guessed password into the victim's real password. In that regard, the list-key turned out to require significantly more changes than the other three methods, making it the least susceptible to shoulder surfing attacks. There are a few possible reasons for that. Firstly, a lot of insertions were required. Participants' guesses of list-key passwords were often very short, most likely because they could not follow the experimenter inputting the password, and gave up. Secondly, the method's input type affected their ability to see and memorize the characters being input. Consequently, more deletions and substitutions were needed for the list-key than for the mouse equivalent of the same method. Finally, since association list characters can be either right or wrong, the wrong guesses have a larger impact on the score than the partially wrong characters in textual and GCPS passwords.

\textit{N-Grams.} Much like the previous metric, N-Grams also indicated the superiority of the list-key method over the other three methods, albeit for a different reason. We observed that it was extremely difficult for the attackers to correctly guess more than just one character at a time. That strengthens our previous assumptions that most of correct guesses in list-key passwords were actually blind. For both graphical passwords, including the list-mouse variant, it was easier for the attackers to observe several consecutive characters at a time. Textual passwords performed only marginally better, mainly because the participants needed only to see one or two characters, before they could attempt to deduce which characters followed. The individual participants taking good guesses increased the score.

\textbf{Guessing Order.} This cluster of metrics evaluates the extent to which the guessed password could be helpful to a malicious observer, particularly when used to increase the efficiency of a brute-force attack. For that, two modified brute-force approaches were considered as the attacker's most likely tactics. Used as a common measure of password strength, entropy was also examined, particularly to gauge the effect of the guess on the decrease of search space, which directly influences brute-force attacks.

Most notably, and not in accordance with the previous two metric clusters, we observed a significant difference between the GCPS and the other three authentication methods. According to the combined set of metrics, the attacker's guess has the potential of significantly reducing the required effort to uncover the victim's chess password. In other words, the correct and partially correct guesses of chess characters account for a noticeable decrease in the size of the search space still necessary to traverse, enabling the attacker to recover the password of equal strength faster. We believe this disparity between the clusters largely depends on the character pool size, which neither characteristics nor distance metrics actively take into account. They are affected by it to some degree, particularly in the sense of guessing probability, but only the guessing metrics consider its effect on password security. The GCPS inflates its theoretical search space by stacking several independent layers of character features: each character is composed of a figure, its color, and its position on the chessboard. Guessing (even just a part of) the character therefore has a much greater impact on the reduction of search space than in other types of authentication methods. 

\textit{Pool Guess*.} That is particularly evident in the case of the pool-based guessing order. Adversaries are the most likely to employ this strategy when they have a high level of confidence in their own guesses. The modified brute-force attack gives priority to the guessed characters, traversing the search space mostly by permuting the characters in the guessed password. Therefore, if every correctly guessed chess character substantially decreases the size of the search space, the correct password will be found much sooner. 

\textit{Position Guess*.} The same conclusion cannot be made for the position-based guessing order. This strategy is more reasonable in situations in which the guesses are more of a guide. Since a greater emphasis is given to the positions of characters within the password rather than the characters themselves, a correct character will only decrease the search space if it was also placed in the right spot. In this situation, authentication methods with smaller character pools are at a disadvantage, not only because there is a higher chance for an attacker to blind-guess the character, but mainly because there is fewer possible substitutions for every individual wrong character. As such, the list method scored worse than for the previous metric, being only marginally better than the GCPS. Instead, textual passwords were shown to be the most resistant to the position-based brute-force approach.

\textit{Entropy*.} Overall, the difference in password strength (as measured by entropy) between the guessed and the actual password was the least prominent in chess passwords, while it was much higher for the remaining three methods. The results are compliant with the pool-guess metric, albeit for another reason. Scores of both the GCPS and the association lists were primarily affected by the length of the guessed password. While most participants failed to provide a guessed list password of sufficient length because they could either not remember it, or gave up half-way, most GCPS guesses were of similar size to the original. On the other hand, textual passwords' entropy score depended more on the character pool size, as most participants' guesses did not include characters from all four independent categories we considered: numeric, lowercase alphabetic, uppercase alphabetic, and symbols. Neither of the two graphical methods could be affected by the pool size, because all components of the password character (e.g. piece, position, and color in GCPS) are always used. 

\subsection{Comparison of active and passive observers}

\subsubsection{Login Times}

When comparing active and passive observers, we found no significant differences in login times for any authentication method except the mouse variation of the association lists (\textit{U} = 887, \textit{p} = .001), for which the active participants (Mdn = 74.51s) were much faster than the passive ones (Mdn = 112.39s). The prevailing differences in login times \textit{between} the methods were influenced by password input times. However, these should be the same for any groups \textit{within} the same method. That means the observed differences in the groups depended on password recall times. 

It is difficult to provide a convincing reasoning as to why the differences between the two types of observers are only prominent for the list-mouse method. Our most viable interpretation relies on the graphical aspect of the method. While the active participants attempted to input the password they deliberately memorized during the shoulder surfing stage, the passive participants likely needed additional time to try and recollect the words they saw on the screen. The same was not necessarily true for the keyboard variation of the list method; both active and passive participants not only had trouble familiarizing themselves with the method, but were equally lost in regards to what they had observed. 

Both groups also struggled with the interface for chess passwords. While active observers tried to recall the password they saw, the passive observers counted on being able to remember the state of the chessboard as it was during the shoulder surfing stage. However, unlike for the list-mouse method, they gave up much faster. Even though both methods are visual, it was clear that the participants recalling the chess passwords felt they had a smaller chance of remembering them. That could be because not every participant was familiar with the game of chess, and was therefore never trained to distinguish between, and recall different states of the chessboard. On the other hand, associations are intuitive, and may to some extent seem logical, which was why some passive observers might have attempted to guess the password even if they didn't recall the words they have seen. Finally, both login times for textual passwords were fairly short, due to the participants' familiarity with the authentication method. Not much time was devoted to password recall in this case. If the participants remembered the observed password, they could type it in, while in the opposite case, it would be just as simple for them to input a blind guess.

\subsubsection{Vulnerability Metrics}

Table \ref{table:3} reports the results of pairwise comparisons between the four authentication methods for active and passive observers, respectively. For the sake of brevity, however, an in-depth analysis has been omitted. Instead, the next section focuses on disparities between the two types of observers for each of the considered methods. The relevant \textit{p}-values are given in the rightmost section of Table \ref{table:2}.

{\setlength\doublerulesep{0.4pt}
     
  \begin{table*}[t]
    \begin{adjustwidth}{-1.4in}{-.5in} 
    \centering
  \begin{tabular}{ lcccccc|ccccccc } 
  \toprule[1pt]\midrule[0.3pt]
    & \multicolumn{6}{c}{\textbf{Active Participants}} & \multicolumn{6}{c}{\textbf{Passive Participants}} \\\midrule
    \textbf{Metric} & \textbf{P-C} & \textbf{P-L\textsubscript{k}} & \textbf{P-L\textsubscript{m}} & \textbf{C-L\textsubscript{k}} & \textbf{C-L\textsubscript{m}} & \textbf{L\textsubscript{k}-L\textsubscript{m}} & \textbf{P-C} & \textbf{P-L\textsubscript{k}} & \textbf{P-L\textsubscript{m}} & \textbf{C-L\textsubscript{k}} & \textbf{C-L\textsubscript{m}} & \textbf{L\textsubscript{k}-L\textsubscript{m}}  \\ \midrule[1pt]
    
    Length Dif & 1 & \sig 0 & \sig 0 & \sig 0 & \sig 0 & 1 & \sig 0 & .384 & 1 & \sig 0 & \sig 0 & 1 \\\midrule
    Same Chars & 1 & .414 & 1 & 1 & 1 & .63 & .186 & .57 & \sig .006 & 1 & 1 & .456 \\
    Correct First & 1 & \sig 0 & 1 & \sig 0 & 1 & \sig 0 & \sig 0 & .132 & 1 & \sig 0 & \sig 0 & .252 \\
    Right Spot & 1 & \sig 0 & \sig .042 & \sig 0 & .114 & \sig 0 & .318 & \sig 0 & \sig .006 & \sig 0 & \sig 0 & .636 \\
    LCS & 1 & \sig 0 & \sig 0 & \sig 0 & 1 & \sig 0 & .252 & 1 & 1 & \sig .006 & \sig .036 & 1 \\
    Dif in Guess* & 1 & \sig 0 & \sig 0 & \sig 0 & \sig 0 & .36 & 1 & \sig 0 & \sig 0 & \sig 0 & \sig 0 & .168 \\
    \textbf{Characteristics} & 1 & .054 & 1 & 1 & 1 & \sig 0 & .246 & 1 & \sig .024 & 1 & 1 & \sig .042 \\\midrule
    
    Jaccard & 1 & 1 & .108 & 1 & .252 & .222 & .204 & \sig .024 & \sig 0 & 1 & .084 & .504 \\
    Jaro-Winkler & 1 & 1 & \sig 0 & 1 & \sig 0 & \sig 0 & .366 & .156 & \sig .036 & 1 & 1 & 1 \\
    Cosine & .228 & 1 & 1 & .846 & \sig .042 & .12 & 1 & .126 & \sig 0 & .66 & \sig .006 & .252 \\
    Levenshtein & 1 & \sig 0 & 1 & .336 & 1 & \sig 0 & 1 & 1 & 1 & .102 & .408 & 1 \\
    N-grams & 1 & \sig .006 & 1 & .156 & 1 & \sig 0 & .51 & .948 & 1 & \sig .03 & .372 & .816 \\
    \textbf{Distance} & 1 & \sig 0 & .288 & 1 & .264 & \sig 0 & .612 & .6 & \sig .018 & 1 & 1 & .27 \\ \midrule
    
    Pool Guess* & .486 & \sig .048 & .354 & .294 & .486 & .228 & 1 & 1 & 1 & .348 & .066 & 1 \\
    Position Guess* & \sig .024 & 1 & \sig .012 & 1 & 1 & \sig 0 & \sig .018 & .216 & .876 & \sig .042 & .096 & 1 \\
    Entropy* & .192 & \sig 0 & \sig 0 & \sig 0 & \sig 0 & 1 & \sig 0 & 1 & 1 & \sig 0 & \sig 0 & 1 \\
    \textbf{Guessing Order*} & \sig .012 & \sig .006 & .084 & \sig 0 & \sig 0 & 1 & \sig 0 & 1 & 1 & \sig 0 & \sig 0 & 1 \\\midrule[0.3pt]\bottomrule[1pt]
  \end{tabular}
  \end{adjustwidth}
  \caption{Bonferroni-corrected \textit{p}-values for pairwise comparisons between all authentication methods, with respect to observer type. All \textit{p}-values, significant at .05 level, are shaded. Metrics marked with an asterisk are complementary.}
  \label{table:3}
  \end{table*}
   }

\medskip

\textbf{Password Characteristics.} The GCPS was the only authentication scheme for which no differences were observed between active and passive participants. In all other cases, participants of the passive group were substantially worse at guessing the observed passwords. Our initial expectation was that larger differences between the groups would have been found for the methods the users were not familiar with. Instead, familiarity with the method did not seem to have a decisive effect. 

Based on the results, we concluded that the visual aspect of the GCPS could have contributed to the passive participants' increased success. Even though they were not initially aware of their objective, the passive participants could still follow the chess pieces being placed onto the chessboard. We believe it could have been easier for these participants to subconsciously memorize parts of the GCPS password in view of the well-documented Picture Superiority Effect in memory literature. Meanwhile, despite offering a possible graphical interaction through the use of a mouse, the association list is still inherently dependent on textual constructs. Furthermore, personal associations provide a memory aid to the user, rather than the attacker. With no graphical stimuli to benefit in encoding and retrieval of password concepts, opportunistic observers would be at a disadvantage against deliberate observers who are actively trying to remember the password. The other considered textual authentication methods follow the same principle.

Perhaps interesting to note is the fact that the \textit{p}-values are consistent across the individual metrics for all authentication methods. The only exception is the proportion of same characters in association lists, for which no difference has been observed between the two types of participants. That is likely due to the method's small character pool, which increases the chance of blind guessing the characters appearing in the password. However, as indicated by the rest of the metrics, that does not necessarily mean that passive observers are better at guessing the observed passwords.

\textbf{Distance Metrics.} Similar findings to the previous cluster of metrics can be reported. Distance metrics focus on measuring the similarity between the passwords, indicating that the guesses made by the passive observers of chess passwords were just as similar to the real passwords, as those made by the active observers. That supports our previous hypothesis that graphical passwords are more susceptible to successful shoulder surfing attacks from opportunistic observers. 

Analogously, no disparity has been observed between the individual distance metrics when evaluating the differences between the two observer types for each authentication method. The consensus strengthens the validity of the results, and decreases the likelihood of a statistical error. Once again, the only exception (Jaccard Index) highly relied on the character pool size. 

\textbf{Guessing Order.} In general, the pairwise comparisons yield complementary results. Passive observers of chess passwords not only managed to substantially decrease the differences in password strength between the target passwords, but were exceedingly successful at reducing the search space, too; using the pool-based approach, they scored even better than their active counterparts. That provides an articulate example illustrating just how negatively a graphical component of a password can affect the method's level of resistance against shoulder surfing attacks.

Overall, textual passwords were the only method for which the composite guessing order metric showed significant differences between the two observer groups. The outcome was mainly affected by the large discrepancies in entropy scores. While active participants managed to guess passwords of length and composition similar to that of the original password, the passive participants were much less successful. Their guesses were of variable length, and often did not include all types of characters appearing in the victim's password, particularly uppercase alphabetics and symbols. This is also in line with our assumption that it is much more difficult for opportunistic observers to subconsciously see and memorize textual constructs, particularly when inputting them could mean merely pressing an additional key on the keyboard. Regardless, neither group's guesses managed to substantially decrease the traversable search space, making it the most resistant to modified brute-force attacks. A closer look shows that larger differences were observed in the pool-based guessing order metric, because active observers were better at discerning the individual characters being used in the password. Both groups appeared equally unable to guess the right positions of the correct characters in the password, however.

The opposite can be said for the association lists. Given the relatively small character pool, both active and passive observers were able to guess a similar number of characters appearing in the password, which attributed to a proportional decrease in the search space. When it comes to placing the correctly guessed characters in the correct order, however, active participants were much better, indicating that they were able to follow the words being input on the screen. As expected, the mouse variation group was much more successful, with the scores comparable to those of the graphical GCPS. On the other hand, the keyboard variation group's results were more dispersed, suggesting the participants were either more, or less successful in their guesses, depending on the concentration and skill of an individual participant. In terms of entropy, association lists scored well mostly because the majority of participants gave up with their guessing before getting to the length of the original password. This tendency was observed for active and passive participants alike. Overall, the differences between the two observer groups were too small to affect the composite guessing order metric.

\section{Discussion}

In the broadest sense, few significant differences in shoulder surfing susceptibility were observed between the considered methods. Nonetheless, all of them indicate that graphical passwords are indeed more vulnerable to observational attacks than their textual counterparts. For instance, the variation of association lists where the password was input using a mouse performed consistently worse than the other three methods across characteristic and distance metrics alike. The guesses made for the GCPS password would have notably decreased the required guessing effort when compared to the other methods. Perhaps the most convincing evidence, however, is the comparison between both variations of the association lists. The main reason why two variations of the same authentication method were considered in the first place was to eliminate the effect of any other possible independent variable on shoulder surfing susceptibility. The only difference between the two groups, the input type, served to illustrate the divergent human-computer interaction that characterizes textual and graphical passwords. In that sense, the graphical variation (i.e. mouse input) was inferior to the textual variation (i.e. keyboard input) in almost all vulnerability metrics. 

It would perhaps be expected for the shoulder surfing susceptibility levels to vary more prominently between the methods, especially given the outlined impact of the textual-graphical disparity on the measure. The reason why that is not the case lies in the complexity of the susceptibility measure. Aside from the aforementioned differences between textual and graphical passwords, the vulnerability to shoulder surfing attacks can also be affected by other factors, such as the mapping between the memorized concept and the encoded password characters, or the scheme's underlying security.

For example, it would be easy to make a false assumption that the method with a greater proportion of guessed characters is also the most vulnerable to shoulder surfing attacks. While it may be true that the attacker might be able to make successful guesses easier, that does not necessarily mean that he or she would have an easier time finding the correct password, thanks to the method's large character pool. For that reason, prior to the beginning of the experiment, significant effort had been devoted toward taking method inequalities into account and normalizing such differences to allow for a fair comparison. As a result, our measure of susceptibility to shoulder surfing attacks did not only encompass how much the attackers were able to see, memorize, and replicate, but also to what extent their guess could aid them in recovering the actual password.

Our experimental design and the inclusion of multiple vulnerability metrics allowed us to identify and assess the factors responsible for the differences in susceptibility levels between the methods. While the graphical component appeared to \textit{increase} the susceptibility score, a larger character pool or a longer password to guess caused the score to \textit{decrease}. The inversely proportional variables ultimately led toward evening the differences in susceptibility levels, until only minor distinctions between the authentication methods could be emphasized. That does not diminish the importance of the results obtained in this study, however. Quite the contrary: the empirical evidence of the graphical passwords' negative effect on the vulnerability to shoulder surfing attacks is an important contribution of our study. 

Furthermore, the only reason why the considered graphical methods could compete with textual in terms of susceptibility to shoulder surfing was because of their mathematical robustness and enhanced security. This is an important implication for the future research in the field of graphical passwords: while researchers should be wary of their inherent predisposition to shoulder surfing attacks, that should not stop them from attempting to devise graphical schemes resistant to shoulder surfing. Depending on its design and other factors, each scheme will be more, or less susceptible to shoulder surfing attacks. Increasing the theoretical (and particularly practical) search space should decrease the chance of such an attack being successful, while also having a positive effect on the scheme's underlying security. On the other hand, it will negatively affect the scheme's usability, as it was briefly seen on the example of login times. Finding the balance between, while striving to improve all aspects (security, usability, and deployability) remains the focal point of password security, including graphical passwords.

To get a clear and comprehensive idea of how susceptible it is against shoulder surfing, we recommend for each new authentication method to be empirically evaluated. Experimental design should depend on the threat model. In our case, a single live observation best served our intention to determine the minimal amount of useful information the attacker can realistically obtain from observing the victim inputting their password. The separation of the participants into opportunistic and deliberate observers allowed for a more fine-grained analysis of the considered authentication methods, shedding light on graphical passwords' liability to circumstantial observations. Future studies should tailor their experimental design according to their requirements and research directions. This paper provides a framework that should help with achieving that goal. 

\subsection{Limitations}

\textbf{Sample.} The participants partaking in this study may not be completely representative of the general population. However, while anyone could execute a shoulder surfing attack, we believe the distribution is skewed towards certain types of attackers. In our study, we modelled a profile of a young, tech-savvy malicious observer. We believe that a typical (under)graduate student fits that archetype well. To fully comply with our threat model, it would have been sensible to use fast-paced video gamers \cite{Kwon}. As active observers, they would have most likely performed better than our participants, particularly after undergoing observational training. Such experimental setup should be explored in a subsequent study. Furthermore, future studies should also examine and compare the success rates of other types of observers (e.g. work colleagues in corporate environments).

\textbf{Data analyses.} The main purpose of the study was to compare the susceptibility to shoulder surfing attacks between textual and graphical passwords. In that sense, any existing graphical authentication scheme could have been chosen. The motivation for our choice of methods was already presented in the preliminaries section of this paper. Nonetheless, future studies should look into comparing all types of graphical authentication methods. All methods considered in our study were recall-based (textual and GCPS passwords were pure recall-based, whereas Association Lists were cued recall-based), which allowed for an easier comparison. However, they provide little cognitive alleviation and often come with considerable overheads. Including recognition-based passwords into the comparison would allow the researchers to examine whether the cognitive advantages they offer to the users also apply to shoulder surfers. Furthermore, future studies should examine methods' resistance against shoulder surfing attacks within several threat models, particularly live versus recorded, and single versus multiple observations. It should also be interesting to explore the effect of environmental factors on the viability of such attacks, such as the type of the device in use, password strength, the victim's input procedure, and the adversary's observation strategy. In relation to the last, the effect of training could also be investigated. Such evaluations are important for furthering our understanding of the existing password mechanisms and their vulnerabilities.

\textbf{Threats of validity.} The original passwords, observed by the participants in the experiment, may not have represented typical strong passwords. We strove to choose passwords of such length and composition so that the underlying security provided would be equal across all authentication methods. However, we could not ensure that the chosen passwords were not more or less intuitive than the average strong password for a given method would be. The participants were also at differing levels of familiarity with the authentication schemes. In particular, all participants were very familiar with textual passwords, while none knew the other methods. For that reason, active participants were educated about the method they were about to observe prior to the beginning of the experiment. The same could not be done for passive participants, as that would have affected the results. The devices used for testing did not belong to the participants, meaning that they may not have had experience using them.

\subsection{Ethical Considerations}

During the pilot study, the experimental procedure was reviewed by the participating researchers to ensure that study participants would be treated fairly. All participants provided consent prior to the beginning of the study, and had an option to opt-out at any point during the experiment. Their identities were obfuscated and were not taken into account during experimental analysis. Furthermore, all passwords used were created for the purpose of the study, and did not protect any real accounts. As such, there are no ethical concerns related to the participants increasing risk to themselves or others by engaging in the role of shoulder surfers. Instead, it can be argued that their participation increased their awareness of possible risks associated with shoulder surfing attacks.

\section{Conclusion}

Shoulder surfing vulnerability is a complex measure, dependent not only on whether a method is textual or graphical, but also on the individual method's design features that influence the attacker's observation strategy. In this paper, we propose a set of metrics, aimed to capture various aspects affecting a method's susceptibility to observational attacks. We adjust the metrics to consider partial guesses, and normalize them based on the original password length. Finally, we combine individual metrics into three clusters: password characteristics, distance metrics, and guessing order. Using this model, we evaluate four authentication methods in two observation scenarios. That allows us to compare the methods from several points of view, including the type of the authentication scheme, the input method, and the intent of the observer. 

The participants were consistently better at guessing association list passwords when they were input using a mouse as opposed to a keyboard. Equally, active participants were more successful than their passive counterparts in most cases. Both findings empirically support the evidence provided in previous studies. However, composite metrics found few differences between the authentication methods. A false conclusion that graphical methods are not any more susceptible to shoulder surfing than textual methods could easily be made. Fortunately, individual metrics allowed for an in-depth analysis of the obtained results. In particular, the attackers observed and memorized more correct characters appearing in correct positions in graphical passwords, and had a stronger degree of certainty in their guesses. Furthermore, their guesses were often partially correct, indicating that the graphical component made it easier for them to retain at least some information about the observed characters. On the other hand, textual passwords' significantly smaller character pool size rendered the method more prone to successful blind-guessing, evening the differences in vulnerability levels.

Based on that, important conclusions can be drawn. Graphical passwords \textit{are} more vulnerable to observational attacks, because the attackers (whether malicious or not) can observe and memorize graphical constructs easier than textual. However, that does not yet guarantee that a shoulder attack on graphical passwords is more successful than on textual passwords by default. Usually, usability shortcomings in graphical passwords enable higher security. In the scope of shoulder surfing, that means the attackers have a much lower chance of correctly guessing a password character they did not manage to observe in the password. Such factors may influence the attacker's success. Previous studies focused only on the attacker's guessing capabilities, and failed to take other contributing factors into account. The purpose of our study was to provide the researchers with a solution that allows for a full-scale shoulder surfing analysis and comparison of any knowledge-based authentication method. The proposed ensemble supports: (1) objective, comparable, and comprehensive evaluation of a method's susceptibility to shoulder surfing, (2) investigation of underlying reasons affecting a method's shoulder surfing vulnerability by considering the scores of individual metrics, and (3) easy modifications by adding or removing individual metrics and applying weights to them. While by no means an absolute measure, the vulnerability ensemble makes an important step toward capturing the multi-faceted nature of shoulder surfing.

Newly proposed authentication methods should consider employing a more objective and systematic approach to shoulder surfing evaluation. Equally, future studies should re-evaluate existing authentication methods, and compare them to other approaches. A somewhat refined version of our model can be used for that purpose. In our study, we have empirically shown that the susceptibility to shoulder surfing is affected by many independent factors. Different authentication schemes might be more or less susceptible to these attacks due to their design. In our case, vastly different authentication schemes share a similar probability of a successful shoulder surfing attack. However, that might not be the case for every novel method. It is therefore crucial for researchers and innovators to understand the importance of unbiased evaluation, and benchmark comparison with traditional, textual passwords. 

\section{Acknowledgements}

The authors acknowledge the financial support from the Slovenian Research Agency (research core funding no. P2-0057).

\bibliographystyle{ACM-Reference-Format}
\bibliography{bibliography}

\clearpage

\begin{appendices}
\section{Literature Review}

{\setlength\doublerulesep{0.4pt}
\setlength\tabcolsep{1.55pt}
  \begin{table*}[!htb]
  \begin{adjustwidth}{-1in}{-.5in}  
  \centering
  \begin{threeparttable}
  \begin{tabular}{lll|cccccccccc|ccccccccc|cccc|ccc } 
  
    & & & \multicolumn{10}{c}{\textbf{Method Design \cite{Schaub13}}} & \multicolumn{16}{c}{\textbf{Experimental Setup}}  \\
    & & & \multicolumn{4}{c}{CHR} & \multicolumn{4}{c}{DSG} & \multicolumn{2}{c}{CPB} & \multicolumn{9}{c}{Shoulder Surfing Exp.} & \multicolumn{4}{c}{Compare} & \multicolumn{3}{c}{Metrics} \\\midrule
    \textbf{Category} & \textbf{Scheme} & {\rotatebox[origin=lc]{90}{\textbf{Reference}}} & {\rotatebox[origin=lc]{90}{\textit{Security}}} & {\rotatebox[origin=lc]{90}{\textit{Observ. Resistance}}} & {\rotatebox[origin=lc]{90}{\textit{Efficiency}}} & {\rotatebox[origin=lc]{90}{\textit{Memorability}}} & {\rotatebox[origin=lc]{90}{\textit{Spatial Arrangement}}} & {\rotatebox[origin=lc]{90}{\textit{Temporal Arrangement}}} & {\rotatebox[origin=lc]{90}{\textit{Visual Cues}}} & {\rotatebox[origin=lc]{90}{\textit{Interaction Method}}} & {\rotatebox[origin=lc]{90}{\textit{Context of Use}}} & {\rotatebox[origin=lc]{90}{\textit{Constraints}}} & {\rotatebox[origin=lc]{90}{\textit{\# Participants}}} & {\rotatebox[origin=lc]{90}{\textit{Participant Profile}}} & {\rotatebox[origin=lc]{90}{\textit{Writing Aid}}} & {\rotatebox[origin=lc]{90}{\textit{Victim/Observer}}} & {\rotatebox[origin=lc]{90}{\textit{Active/Passive}}} & {\rotatebox[origin=lc]{90}{\textit{Live/Video}}} & {\rotatebox[origin=lc]{90}{\textit{Positioning}}} & {\rotatebox[origin=lc]{90}{\textit{\# Observations}}} & {\rotatebox[origin=lc]{90}{\textit{\# Passwords}}} &
    {\rotatebox[origin=lc]{90}{\textit{w/ Input Type}}} &{\rotatebox[origin=lc]{90}{\textit{w/ Other Groups}}} & {\rotatebox[origin=lc]{90}{\textit{w/ Other Schemes}}} & {\rotatebox[origin=lc]{90}{\textit{w/ Textual Passwords}}} & {\rotatebox[origin=lc]{90}{\textit{Guessing Accuracy (\%)}}} & {\rotatebox[origin=lc]{90}{\textit{\# Observations}}} & {\rotatebox[origin=lc]{90}{\textit{Distance Metrics}}} \\ \midrule[1pt]

     Various & Various & \cite{Schaub13} & \hatchE{$\circ$} & \hatchE{} & \hatchB{$\bullet$} & \hatchE{$\circ$} & \hatchW{$\circ$} & \hatchE{$\circ$} & \hatchW{} & \hatchW{$\circ$} & \hatchW{} & \hatchE{} & \hatchW{$\circ$} & \hatchE{} & \hatchB{$\bullet$} & \hatchE{$\circ$} & \hatchW{} & \hatchB{$\bullet$} & \hatchB{$\bullet$} & \hatchE{} & \hatchE{$\circ$} & \hatchW{} & \hatchE{$\bullet$} & \hatchE{$\bullet$} & \hatchW{} & \hatchW{$\circ$} & \hatchE{} & \hatchW{} \\\midrule
    Recognition & Doodles & \cite{Jebriel} & \hatchW{} & \hatchE{} & \hatchE{$\circ$} & \hatchW{$\circ$} & \hatchE{$\bullet$} & \hatchE{$\circ$} & \hatchW{} & \hatchE{$\bullet$} & \hatchW{} & \hatchE{} & \hatchW{$\circ$} & \hatchE{$\circ$} & \hatchB{$\bullet$} & \hatchB{$\bullet$} & \hatchW{} & \hatchE{$\circ$} & \hatchE{$\circ$} & \hatchE{} & \hatchE{$\circ$} & \hatchE{$\bullet$} & \hatchE{$\bullet$} & \hatchW{} & \hatchW{} & \hatchE{$\bullet$} & \hatchE{} & \hatchW{} \\
    & EvoPass & \cite{Yu} & \hatchW{} & \hatchB{$\circ$} & \hatchB{$\bullet$} & \hatchE{$\bullet$} & \hatchW{$\circ$} & \hatchB{$\bullet$} & \hatchE{$\circ$} & \hatchW{} & \hatchB{$\bullet$} & \hatchE{} & \hatchW{} & \hatchW{} & \hatchE{} & \hatchE{$\circ$} & \hatchW{} & \hatchE{$\circ$} & \hatchE{$\circ$} & \hatchB{$\bullet$} & \hatchB{$\bullet$} & \hatchW{} & \hatchE{$\bullet$} & \hatchE{$\bullet$} & \hatchW{} & \hatchE{$\bullet$} & \hatchB{$\bullet$} & \hatchW{} \\
    & PassImages & \cite{Dunphy} & \hatchW{} & \hatchE{} & \hatchE{$\circ$} & \hatchW{$\circ$} & \hatchW{$\circ$} & \hatchB{$\bullet$} & \hatchW{} & \hatchW{} & \hatchW{} & \hatchE{} & \hatchW{} & \hatchE{$\circ$} & \hatchE{} & \hatchB{$\bullet$} & \hatchW{} & \hatchE{$\circ$} & \hatchE{$\circ$} & \hatchB{$\bullet$} & \hatchE{$\circ$} & \hatchW{} & \hatchE{$\bullet$} & \hatchW{} & \hatchW{} & \hatchW{} & \hatchB{$\bullet$} & \hatchW{} \\
    & IllusionPIN & \cite{Papadopoulos} & \hatchW{} & \hatchB{$\bullet$} & \hatchE{$\circ$} & \hatchE{$\bullet$} & \hatchW{$\circ$} & \hatchB{$\bullet$} & \hatchE{$\circ$} & \hatchW{} & \hatchB{$\bullet$} & \hatchE{} & \hatchW{} & \hatchE{$\circ$} & \hatchE{} & \hatchB{$\bullet$} & \hatchW{} & \hatchE{$\circ$} & \hatchW{} & \hatchB{$\circ$} & \hatchE{$\circ$} & \hatchW{} & \hatchE{$\bullet$} & \hatchW{} & \hatchW{} & \hatchE{$\bullet$} & \hatchE{} & \hatchW{} \\
    & PassFaces & \cite{Tari} & \hatchW{} & \hatchE{} & \hatchE{$\circ$} & \hatchE{$\bullet$} & \hatchW{$\circ$} & \hatchE{$\circ$} & \hatchW{} & \hatchE{$\bullet$} & \hatchW{} & \hatchE{} & \hatchW{} & \hatchE{$\circ$} & \hatchB{$\bullet$} & \hatchE{$\circ$} & \hatchW{} & \hatchE{$\circ$} & \hatchB{$\bullet$} & \hatchE{} & \hatchE{$\circ$} & \hatchE{$\bullet$} & \hatchW{} & \hatchE{$\bullet$} & \hatchE{$\bullet$} & \hatchE{$\bullet$} & \hatchE{} & \hatchW{} \\
    & Various & \cite{Cain17} & \hatchW{} & \hatchE{} & \hatchE{$\circ$} & \hatchE{$\bullet$} & \hatchE{$\bullet$} & \hatchB{$\bullet$} & \hatchE{$\circ$} & \hatchW{$\circ$} & \hatchE{$\circ$} & \hatchE{} & \hatchW{} & \hatchE{$\circ$} & \hatchE{} & \hatchE{$\circ$} & \hatchW{} & \hatchW{} & \hatchW{} & \hatchB{$\circ$} & \hatchE{$\circ$} & \hatchW{} & \hatchW{} & \hatchE{$\bullet$} & \hatchW{} & \hatchE{$\bullet$} & \hatchE{} & \hatchW{} \\\midrule
    Pure-Recall & DAS & \cite{Zakaria} & \hatchE{$\circ$} & \hatchE{} & \hatchE{$\circ$} & \hatchW{$\circ$} & \hatchW{} & \hatchW{} & \hatchB{$\bullet$} & \hatchW{$\circ$} & \hatchW{} & \hatchB{$\circ$} & \hatchW{$\circ$} & \hatchE{$\circ$} & \hatchB{$\bullet$} & \hatchE{$\circ$} & \hatchW{} & \hatchE{$\circ$} & \hatchB{$\bullet$} & \hatchE{} & \hatchE{$\circ$} & \hatchW{} & \hatchE{$\bullet$} & \hatchW{} & \hatchW{} & \hatchE{$\bullet$} & \hatchE{} & \hatchW{} \\
    & Swipe & \cite{Cain16} & \hatchW{} & \hatchE{} & \hatchB{$\bullet$} & \hatchE{$\bullet$} & \hatchW{} & \hatchW{} & \hatchB{$\bullet$} & \hatchW{} & \hatchW{} & \hatchB{$\circ$} & \hatchW{} & \hatchE{$\circ$} & \hatchE{} & \hatchE{$\circ$} & \hatchW{} & \hatchW{} & \hatchW{} & \hatchE{} & \hatchE{$\circ$} & \hatchW{} & \hatchE{$\bullet$} & \hatchW{} & \hatchW{} & \hatchW{$\circ$} & \hatchE{} & \hatchW{} \\
    & SwiPIN & \cite{Wiese16} & \hatchW{} & \hatchE{} & \hatchB{$\bullet$} & \hatchE{$\bullet$} & \hatchW{$\circ$} & \hatchB{$\bullet$} & \hatchB{$\bullet$} & \hatchW{} & \hatchE{$\circ$} & \hatchE{} & \hatchE{$\bullet$} & \hatchB{$\bullet$} & \hatchE{} & \hatchE{$\circ$} & \hatchW{} & \hatchW{} & \hatchW{} & \hatchE{} & \hatchB{$\bullet$} & \hatchW{} & \hatchE{$\bullet$} & \hatchE{$\bullet$} & \hatchW{} & \hatchW{$\circ$} & \hatchB{$\bullet$} & \hatchW{}  \\
    & Bend PWs & \cite{Maqsood} & \hatchW{} & \hatchE{} & \hatchE{$\circ$} & \hatchW{$\circ$} & \hatchW{$\circ$} & \hatchW{} & \hatchB{$\bullet$} & \hatchW{} & \hatchE{$\circ$} & \hatchB{$\circ$} & \hatchW{} & \hatchW{} & \hatchB{$\bullet$} & \hatchE{$\circ$} & \hatchW{} & \hatchE{$\circ$} & \hatchB{$\bullet$} & \hatchE{} & \hatchB{$\bullet$} & \hatchW{} & \hatchE{$\bullet$} & \hatchE{$\bullet$} & \hatchW{} & \hatchW{$\circ$} & \hatchE{} & \hatchW{$\circ$} \\
    & PIN & \cite{Aviv} & \hatchW{} & \hatchE{} & \hatchB{$\bullet$} & \hatchE{$\bullet$} & \hatchW{} & \hatchW{} & \hatchB{$\bullet$} & \hatchW{} & \hatchW{} & \hatchB{$\circ$} & \hatchE{$\bullet$} & \hatchB{$\bullet$} & \hatchE{} & \hatchE{$\circ$} & \hatchW{} & \hatchW{} & \hatchW{} & \hatchB{$\circ$} & \hatchB{$\bullet$} & \hatchE{$\bullet$} & \hatchE{$\bullet$} & \hatchE{$\bullet$} & \hatchW{} & \hatchW{$\circ$} & \hatchE{} & \hatchW{} \\
    & BW-method & \cite{Kwon} & \hatchW{} & \hatchB{$\circ$} & \hatchE{$\circ$} & \hatchE{$\bullet$} & \hatchW{$\circ$} & \hatchB{$\bullet$} & \hatchB{$\bullet$} & \hatchW{} & \hatchE{$\circ$} & \hatchE{} & \hatchW{} & \hatchE{$\circ$} & \hatchE{} & \hatchE{$\circ$} & \hatchW{} & \hatchW{} & \hatchW{} & \hatchE{} & \hatchB{$\bullet$} & \hatchW{} & \hatchW{} & \hatchE{$\bullet$} & \hatchW{} & \hatchE{$\bullet$} & \hatchE{} & \hatchW{} \\ \midrule
    Cued-Recall & Conceal PWs & \cite{Ho} & \hatchW{} & \hatchB{$\circ$} & \hatchE{$\circ$} & \hatchW{$\circ$} & \hatchE{$\bullet$} & \hatchE{$\circ$} & \hatchE{$\circ$} & \hatchW{$\circ$} & \hatchB{$\bullet$} & \hatchE{} & \hatchW{$\circ$} & \hatchW{} & \hatchE{} & \hatchE{$\circ$} & \hatchW{} & \hatchW{} & \hatchW{} & \hatchB{$\bullet$} & \hatchW{} & \hatchW{} & \hatchW{} & \hatchW{} & \hatchW{} & \hatchW{$\circ$} & \hatchE{} & \hatchW{} \\
    & Assoc. Lists & & \hatchE{$\circ$} & \hatchE{} & \hatchE{$\circ$} & \hatchE{$\bullet$} & \hatchE{$\bullet$} & \hatchE{$\circ$} & \hatchE{$\circ$} & \hatchE{$\bullet$} & \hatchE{$\circ$} & \hatchE{} & \hatchE{$\bullet$} & \hatchE{$\circ$} & \hatchE{} & \hatchE{$\circ$} & \hatchE{$\bullet$} & \hatchE{$\circ$} & \hatchE{$\circ$} & \hatchE{} & \hatchE{$\circ$} & \hatchE{$\bullet$} & \hatchE{$\bullet$} & \hatchE{$\bullet$} & \hatchE{$\bullet$} & \hatchE{$\bullet$} & \hatchE{} & \hatchE{$\bullet$}
     \\\midrule[0.3pt]\bottomrule[1pt]
  \end{tabular}
  \begin{tablenotes}\footnotesize
    \item $\bullet$ = meets criteria; $\circ$ = partially meets criteria; \textit{no circle} = does not meet criteria.
    \item \hatchB{} = better than our method design \& experimental setup; \hatchW{} = worse than our method design \& experimental setup; \textit{no background pattern} = no change.
    \item All empirical shoulder surfing studies detailed in the literature review are grouped into the standard graphical password categories. Points are awarded to each scheme for various attributes describing the scheme's design and the shoulder surfing experiment conducted. Attributes describing the method's design were originally proposed by Schaub et al. \cite{Schaub13}. All considered attributes and the criteria for awarding points are described on the following page.
    \end{tablenotes}
  \end{threeparttable}
  \end{adjustwidth}
  \caption{Comparison of shoulder surfing experiments across the various authentication schemes}
  \label{table:0}
  
  \end{table*}
}

\clearpage
\footnotesize

\subsection{Method Design}
    \subsubsection{Password Characteristics (CHR)}
    
    \begin{itemize}[label={},leftmargin=15pt]
        \item \textbf{Security.} ($\bullet$) Theoretical password space for typical passwords is sufficient against brute-force attacks ($> 10^{20}$); ($\circ$) brute-force attacks against typical passwords are possible, but unlikely ($> 10^{17}$); ( ) typical passwords are not safe against brute-force attacks ($< 10^{17}$)
        \item \textbf{Observation Resistance.} ($\bullet$) Scheme is entirely resistant to video observation; ($\circ$) scheme is partially resistant to video observation (e.g. multiple observations required); \linebreak ( ) scheme is not resistant to video observation;
        \item \textbf{Efficiency.} ($\bullet$) Typical login time is comparable to textual passwords ($< 10s$); ($\circ$) typical login time is still reasonable for users ($< 60s$); ( ) typical login time is unreasonable ($> 1m$);
        \item \textbf{Memorability.} ($\bullet$) Users can consistently remember their passwords ($> 90\%$); ($\circ$) users can remember their passwords most of the time ($> 70\%$); ( ) users have trouble recalling their passwords ($< 70\%$);
    \end{itemize}
    
    \subsubsection{Design Features (DSG)}
    
    \begin{itemize}[label={},leftmargin=15pt]
        \item \textbf{Spatial Arrangement.} ($\bullet$) Visual elements are randomized and there are many of them; ($\circ$) there are a few randomized visual elements or there are many fixed elements; ( ) the scheme has a few fixed elements;
        \item \textbf{Temporal Arrangement.} ($\bullet$) The scheme provides multiple challenge rounds with changing cues; ($\circ$) authentication requires multiple challenges on a fixed background; ( ) the scheme assumes a single challenge;
        \item \textbf{Visual Cues.} ($\bullet$) There are no visual cues; ($\circ$) visual cues are small or have a low level of detail; ( ) visual cues are large or have a high level of detail;
        \item \textbf{Interaction Method.} ($\bullet$) The password can be input through a keyboard; ($\circ$) the password can be input with a mouse; ( ) the password can be input with a finger;
    \end{itemize}

    \subsubsection{Device Capabilities (CPB)}
    
    \begin{itemize}[label={},leftmargin=15pt]
        \item \textbf{Context of Use.} ($\bullet$) There is a low probability of an observer guessing a password even after multiple observations; ($\circ$) the observer can potentially guess a password after multiple observations; ( ) the observer might guess a password after a single observation;
        \item \textbf{Constraints.} ($\bullet$) Authentication can easily be performed despite motoric and vision impairments; ($\circ$) vision-impaired or color blind users can easily authenticate; ( ) users with motoric or vision impairment cannot easily authenticate;
    \end{itemize}
    
\subsection{Experimental Setup}
    \subsubsection{Shoulder Surfing Experiment}
    
    \begin{itemize}[label={},leftmargin=15pt]
        \item \textbf{\# Participants.} ($\bullet$) The shoulder surfing experiment contained at least 100 participants; ($\circ$) the experiment contained at least 30 participants; ( ) the experiment contained less than 30 participants;
        \item \textbf{Participant Profile.} ($\bullet$) The participants represent the general population; ($\circ$) most participants are university students; ( ) profile of participants was not given;
        \item \textbf{Writing Aid.} ($\bullet$) The participants were allowed to record their observations (e.g. using a pen and paper); ( ) participants did not record their observations;
        \item \textbf{Victim/Observer.} ($\bullet$) The participants played both the roles of victims and observers; ($\circ$) participants acted as observers; ( ) participants were in the role of a victim;
        \item \textbf{Active/Passive.} ($\bullet$) The participants acted as active and passive observers; ($\circ$) participants were cast in the role of active observers; ( ) participants were passive observers;
        \item \textbf{Live/Video.} ($\bullet$) Both live and video observations were included; ($\circ$) the experiment consisted of live observations; ( ) the experiment included only video observations;
        \item \textbf{Positioning.} ($\bullet$) The participants were able to choose their positioning; ($\circ$) the participants had a fixed positioning, but were allowed to assume a comfortable position; ( ) the participants had a strictly fixed positioning;
        \item \textbf{\# Observations.} ($\bullet$) Unlimited number of observations was allowed for each password; ($\circ$) the participants had multiple observations to guess a password; ( ) the participants had a single observation to guess a password;
        \item \textbf{\# Passwords.} ($\bullet$) The participants had to guess several passwords for several schemes; ($\circ$) the participants were required to guess several passwords for one scheme or one password for several schemes; ( ) the participants guessed one password for a single scheme;
    \end{itemize}
    
    \subsubsection{Comparability}
    
    \begin{itemize}[label={},leftmargin=15pt]
        \item \textbf{/w Input Type.} ($\bullet$) The experiment includes comparison between multiple input types (e.g. keyboard vs mouse); ( ) the experiment does not include comparison between input types;
        \item \textbf{/w Other Groups.} ($\bullet$) The study includes different configurations of shoulder surfing a particular scheme (e.g. password strength, viewing angles, device, etc.); ( ) no comparisons between different shoulder surfing setups are conducted;
        \item \textbf{/w Other Schemes.} ($\bullet$) The considered scheme is compared to other schemes in terms of shoulder surfing vulnerability; ( ) there are no comparisons with other authentication schemes;
        \item \textbf{/w Textual Passwords.} ($\bullet$) Comparison of shoulder surfing vulnerability with textual passwords is included; ( ) the considered scheme is not compared to textual passwords;
    \end{itemize}
    
    \subsubsection{Metrics}
    
    \begin{itemize}[label={},leftmargin=15pt]
        \item \textbf{Guessing Accuracy.} ($\bullet$) The percentage of the password being guessed correctly is measured; ($\circ$) binary measure of whether the password was correctly guessed or not is employed; the guessing accuracy is not measured;
        \item \textbf{\# Observations.} ($\bullet$) Guessing accuracies for all observations are recorded; number of observations necessary to correctly guess the password is reported; ( ) number of observations necessary to guess the password is not given;
        \item \textbf{Distance Metrics.} ($\bullet$) Multiple distance metrics are measured to determine similarity between the correct and guessed passwords; ($\circ$) a single distance metric is employed (e.g. Levenshtein); ( ) no distance metrics are measured;
    \end{itemize}

\end{appendices}

\end{document}